\documentclass{pasj01}

\usepackage{color}
\usepackage{mathrsfs}

\begin{document}

\author{Mariko \textsc{Kimura}\altaffilmark{1,*},
        Yoji \textsc{Osaki}\altaffilmark{2},
        Taichi \textsc{Kato}\altaffilmark{1},
        and Shin \textsc{Mineshige}\altaffilmark{1}
        }
\email{mkimura@kusastro.kyoto-u.ac.jp}

\altaffiltext{1}{Department of Astronomy, Graduate School of Science, Kyoto University, Oiwakecho, Kitashirakawa, Sakyo-ku, Kyoto 606-8502}
\altaffiltext{2}{Department of Astronomy, School of Science, University of Tokyo, Hongo, Tokyo 113-0033}

\title{Thermal-viscous instability in tilted accretion disks: a possible application to IW And-type dwarf novae}

\Received{} \Accepted{}


\SetRunningHead{Kimura et al.}{}

\maketitle

\begin{abstract}

IW And stars are a subgroup of dwarf novae characterized 
by repetitive light variations of the intermediate-brightness 
state with oscillations, which is terminated by brightening.
This group of dwarf novae is also known to exhibit a wide 
variety even within one system in long-term light curves including 
usual dwarf-nova outbursts, Z Cam-type standstills, and so on, 
besides the typical IW And-type variations mentioned above.
Following the recent observations suggesting that some 
IW And stars seem to have tilted disks, we have investigated 
how the thermal-viscous instability works in tilted accretion 
disks in dwarf novae and whether it could reproduce the essential 
features of the light curves in IW And stars.
By adopting various simplifying assumptions for tilted disks,
we have performed time-dependent one-dimensional numerical 
simulations of a viscous disk by taking into account various 
mass supply patterns to the disk; that is, the gas stream from 
the secondary star flows not only to the outer edge of the disk 
but also to the inner portions of the disk.
We find that tilted disks can achieve a new kind of accretion 
cycle, in which the inner disk almost always stays in the hot state 
while the outer disk repeats outbursts, thereby reproducing 
alternating mid-brightness interval sometimes with dips and 
brightening, which are quite reminiscent of the most characteristic 
observational light variations of IW And stars.  
Further, we have found that our simulations produce diverse
light variations, depending on different mass supply patterns
even without time variations in mass transfer rates.  This could
explain the wide variety in long-term light curves of IW And stars.  

\end{abstract}

\section{Introduction}

Cataclysmic Variables (CVs) are close binary systems 
composed of a white dwarf (the primary) and a low-mass 
cool star filling the critical Roche lobe (the secondary).  
An accretion disk is formed around the primary white dwarf 
via Roche-lobe overflow from the secondary in non-magnetic CVs.  
Dwarf novae (DNe), one subclass of CVs, exhibit 
transient events called ``outbursts'' with amplitudes of 
2--6 mag with a typical timescale of several 10 days 
due to sudden brightening of accretion disks 
(\cite{war95book} for a general review).  

Thermal limit-cycle instability explains normal dwarf-nova 
outbursts (\cite{osa96review} for a review).  
Partial ionization of hydrogen triggers thermal instability 
in a disk \citep{hos79DImodel}, and the disk shows 
bi-stable states: the hot state with a high accretion rate 
and the cool state with a low accretion rate.  
The disk jumps between the two stable states, because 
they sandwich an unstable state on the thermal 
equilibrium curve, which appears due to significant effect 
of convection \citep{mey81DNoutburst}.  
This limit cycle drastically increases/decreases the mass 
accretion rate onto the primary white dwarf.  
If the mass transfer rate from the secondary star is less 
than the minimum rate for keeping the entire disk hot, which 
is denoted as $\dot{M}_{\rm crit}$, the systems experience 
sporadic outbursts.  
If the mass transfer rate is higher than $\dot{M}_{\rm crit}$, 
the systems show constant high disk luminosity, and are 
called nova-like stars (NLs) \citep{sma83DN}.  

However, we cannot understand several kinds of dwarf-nova 
outbursts only by the simple disk instability model with 
constant mass transfer rates from the secondary.  
One group of CVs called ``Z Cam-type DNe'' are believed 
to be the intermediate systems between DNe and NLs, 
since they show occasionally ``standstills'' having constant 
luminosity intermediate between outburst maxima and minima.  
It is believed that the variations of mass transfer rates 
are necessary to reproduce this kind of phenomena (e.g., 
\cite{mey83zcam}).  
There is at least one attempt to reproduce them by 
the fluctuations of viscosity instead of those of transfer 
rates, but they failed to generate long-lasting standstills 
and several consecutive outbursts \citep{ros17zcam}.  

\citet{sim11zcamcamp1} noticed the presence of two unusual
DNe which show repeated outbursts to standstills terminated 
by brightening (instead of fading in ordinary Z Cam stars).  
After that, \citet{szk13iwandv513cas} called these objects 
``anomalous Z Cam stars'' and discussed the potential relation
with small outbursts in NLs.  
Recently, \citet{kat19iwand} found three more objects showing 
the same behavior as that reported by these authors and named 
these objects ``IW And-type stars'', and has pointed out that 
more or less regular repetition of 
standstills terminated by brightening are common to these 
objects (see also Sec.~2).  
He further suggested the presence of a previously unknown 
type of limit-cycle oscillation in IW And-type stars.  

\citet{ham14zcam} explored for the first time the cause of 
``anomalous Z Cam phenomenon'' and proposed a model in which 
the variation in mass transfer rates from the secondary star 
is responsible for that phenomenon.  However, there is 
no positive evidence for triggering modulations of transfer 
rates, and also, the brightening of bright spots, which should 
occur simultaneously with the sudden increase of transfer rates,  
are observationally undetected \citep{hon01nloutburst,sch19accnc}.  
\citet{ham14zcam} have sought the cause of 
the light variations of IW And-type stars 
in the outside of the disk, i.e., enhanced mass transfer 
from the secondary star.  However, it would be more preferable 
if a previously unknown type of limit-cycle oscillation is found 
within the disk, as suggested by \citet{kat19iwand}.  
Since the standard thermal-viscous instability model is very 
unlikely to produce 
standstills terminated by brightening, we need to seek some new 
aspects not considered in the standard disk instability model.  
To this end, recently, time-resolved optical photometry gave us 
a clue, the detections of negative superhumps in some of 
IW And stars.  

\citet{gie13j1922} and \citet{arm13aqmenimeri} observed negative 
superhumps in some DNe that were identified to 
be IW And-type stars later (e.g., \cite{kat19iwand}).  
The negative superhumps are photometric light modulations 
with periods slightly shorter than the orbital period, 
i.e., they show the negative excess to the orbital period 
\citep{har95v503cyg}, and they are interpreted as the transit 
of the bright spot on the tilted and/or warped disk misaligned 
to the orbital plane, which experiences retrograde nodal 
precession \citep{woo00SH,mur02warpeddisk,woo07negSH}.  
Importantly, this interpretation means that the gas stream 
often flows into the inner part in tilted disks, while it 
always collides with the outer edge of the disk 
in non-tilted disks.  
In fact, \citet{kat19iwand} has suggested that standstills 
in IW And-type DNe may somehow be maintained as the inner part 
of the disk stays in the hot state, while the outer part 
is in the cool state, and that the thermal instability 
starting from the outer part of the disk terminates 
standstills.  
The mass input in the tilted disk might achieve 
such a new limit cycle by keeping the inner disk hot 
during standstills in IW And-type DNe.  

Motivated by this suggestion, we study the disk instability 
model in the case of tilted accretion disks.  
The aim of the present study is twofold: one is to 
investigate how the thermal instability works in tilted 
accretion disks and another is to see to what extent 
\textcolor{black}{the thermal instability in the tilted disk} 
could explain the essential features of light variations in 
IW And-type DNe.  
In Sec.~2, we introduce the observational light variations 
in IW And-type stars and specify their essential features, 
which we aim at reproducing. 
In Sec.~3, we give our assumptions and the method of our 
simulations in time evolution of the one-dimensional disk.  
In Sec.~4, we calculate the patterns of mass input in tilted 
disks, a key of our simulations, and list the calculated 
models.  In Sec.~5, we present the numerical simulations.  
We discuss our results in Sec.~6, comparing them with observations.  
Finally, we give our summary in Sec.~7.

\section{Observational light variations in IW And-type stars}

In these couple of years, we have obtained much more knowledge 
about the light variations in IW And stars than we did when 
the first example of the IW And-type DNe was recognized.  
Here we introduce briefly what the essential features of 
the IW And-type light variations are on the basis \textcolor{black}{of} 
the latest observations.  

\begin{figure}[htb]
\begin{center}
\FigureFile(80mm, 50mm){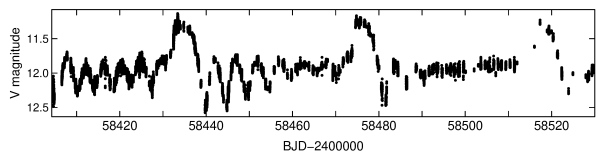}
\end{center}
\caption{An example of the IW And-type phenomenon from a part of the 2018 light curves of IM Eri, which are obtained by a campaign led by Variable Star Network (VSNET).  All of the IW And-type phenomenon of this object in 2018 is presented in \citet{kat19imeri}.  Here, BJD is barycentric Julian date.}
\label{imeri}
\end{figure}

\begin{figure}[htb]
\begin{center}
\FigureFile(80mm, 50mm){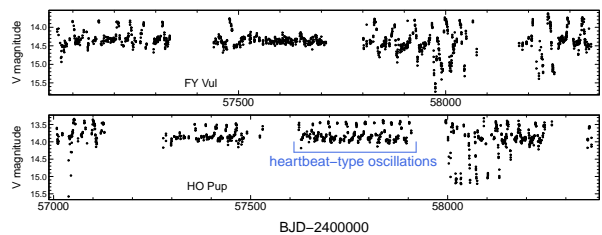}
\end{center}
\caption{A wide variety in long-term light curves of FY Vul and HO Pup, IW And-type stars.  We have obtained the data from ASAS-SN data archive \citep{dav15ASASSNCVAAS}.  }
\label{obsLCs}
\end{figure}

IW And-type stars commonly show repetitive 
light variations as pointed out in \citet{kat19iwand}.  
Figure \ref{imeri} displays the typical IW And-type 
light variations of IM Eri\footnote{\citet{kat19imeri} 
report in detail this kind of light variations of this object, 
which was observed in 2018.}, which was identified as 
an IW And-type star by \citet{kat19iwand}.  
The most characteristic feature of the typical IW And-type 
light variations is ``quasi-standstills'' terminated 
by brightening, and they are one of the essential 
features of light variations in IW And stars.  
Here quasi-standstills are the state in 
intermediate brightness with (damping) oscillatory 
variations, which are different from standstills 
with almost constant luminosity in \textcolor{black}{normal} Z Cam stars.  
We call this the IW And-type phenomenon hereafter.  
Deep luminosity dips occasionally follow brightening 
(see also Figure \ref{obsLCs}).  
The amplitudes of brightening are typically less than 
1 mag.  The averaged interval between brightening 
is $\sim$50 days, but its length never remains in one 
constant value within one object.  

Moreover, long-term photometric surveys for optical transients 
revealed that diverse light variations can be observed 
on long timescales even within one IW And star.  
Figure \ref{obsLCs} illustrates long-term light curves of 
two IW And stars, FY Vul and HO Pup, which were recognized as 
IW And-type DNe by one of the authors 
(T.K.~2018, vsnet-chat 8101\footnote{$<$http://ooruri.kusastro.kyoto-u.ac.jp/mailarchive/vsnet-chat/8101$>$}, 8162\footnote{$<$http://ooruri.kusastro.kyoto-u.ac.jp/mailarchive/vsnet-chat/8162$>$}).  
It is another essential feature of the light variations 
in IW And stars that they would alternate within one object 
between the IW And-type phenomenon, Z Cam-type standstills, 
normal dwarf-nova outbursts, and heartbeat-type oscillations 
on timescales of $\sim$100--1000 days.  
The averaged optical luminosities among those different 
types of light variations are almost constant, and thus 
the time variations in mass transfer rates are unlikely 
for the cause of their diversity.  
Here heartbeat-type oscillations indicated in the lower 
panel of Figure \ref{obsLCs} seem to be a new-type light 
variability, but may be common in IW And stars, because 
similar behavior has been detected also in other objects 
(e.g., \cite{ram16kic920}).  
\textcolor{black}{We use the term ``heartbeat-type oscillations'' 
here, since this type of variability resembles ``heartbeat 
oscillations'' observed in GRS 1915$+$105, a famous black-hole binary, 
which are considered to be caused by limit-cycle accretion triggered 
by the radiation-pressure instability 
\citep{fen04grs1915,wat03instability,nei11grs1915}.  
(This term should not be confused with the pulsation phenomenon of 
``heartbeat stars'' discovered by Kepler data \citep{tho12DTDKepler}.)
}

\section{Method of numerical calculations for time-dependent disks}

\subsection{Basic assumptions of our model}

In this study, we consider how the problem of thermal-viscous 
disk instability in CVs could be applied to the tilted disks.  
Generally speaking, if the disk is tilted, it would have warped 
structures and it would experience retrograde nodal precessions. 
In the binary frame of reference, the disk structure would vary 
with the period of negative superhumps, i.e., the synodic period 
between the orbital period and the precession period. 
To study such complicated problems, we have to perform a full 
3-dimensional hydrodynamical simulations of a non-axisymmetric 
disk whose structures change with time, which is not an easy task.  

However, we think that one of the most important effects of 
the disk tilt on the problem of thermal-viscous disk 
instability is that the gas stream from the secondary star will 
penetrate deeply into the inner part of disk, in contrast to 
the standard case of a non-tilted disk where the gas stream 
arrives mostly in the outer disk edge.
In this paper, we concentrate ourselves to study this particular 
effect of the tilted disk.   
For this purpose, we adopt various simplifying assumptions 
as far as possible.  
We first assume that the tilted disk is not warped and rigidly 
tilted and we take this tilted plane misaligned to 
the orbital plane as our frame of reference.  We further assume 
that the disk is axi-symmetric in this frame of reference 
and we choose the cylindrical coordinates, $r$-$z$ in this plane, 
where $r$ is the distance from the central white dwarf and 
$z$ is the distance perpendicular to this plane.  
By these simplified assumptions our problem reduces to 
the same problem as the standard case of a non-tilted disk, 
and the effects of tilt enter only through the mass supply 
pattern different to that in the standard case.  
In fact, the standard case of a non-tilted disk is regarded 
as one of our special cases with the zero tilt angle. 

Furthermore, we do not discuss any problems of time variations 
shorter than the negative-superhump period and we average 
the mass supply pattern over that period 
(see also \textcolor{black}{Sec.~4.1}).  
This is because we focus on the light 
variations on timescales of days much longer than the period 
of negative superhumps.

\subsection{Basic equations for a viscous disk}

We calculate the time evolution of a geometrically-thin 
and axi-symmetric accretion disk with an assumption of 
a co-planar but tilted disk, basically in the same way 
as in \citet{ich92diskradius}.  
It is noted here that since the original 
computer code by \citet{ich92diskradius} does not exist 
any more, we have newly written our simulation code from 
the beginning in this study, following that paper.

We assume that the disk in our consideration 
is geometrically thin and that it is in hydrostatic equilibrium 
in the vertical $z$-direction.  We adopt the one-zone model 
in the $z$-direction. 
Here, we use the cylindrical coordinates, $(r, \phi, z)$,
with $z$-axis being the disk rotation axis.  
The surface density, $\Sigma$, is then given in the one-zone model by 
\begin{equation}
\Sigma = \int_{-\infty}^{\infty} \rho~dz = 2 \rho_{\rm c} H,  
\label{surface-density}
\end{equation}
where $\rho$ is the density, $H$ is the scale height, 
i.e., the half thickness of the disk which is given 
later by equation (\ref{thickness}), and $\rho_{\rm c}$ is 
the density at the disk mid-plane, respectively.  
Hereafter we express the variables at the disk 
mid-plane with the subscript ``c''.  
We use vertically-integrated basic equations for a viscous disk: 
the equations for conservation of mass, angular momentum, 
and energy, by adopting vertically the one-zone model.  

The equation for mass conservation is 
\begin{equation}
\frac{\partial (2 \pi r \Sigma)}{\partial t} = \frac{\partial \dot{M}}{\partial r} + s, 
\label{mass}
\end{equation}
where 
$\Sigma (r, t)$ is the surface density at a radial 
position, $r$, from the central white dwarf and time, $t$, 
in the unit of g~cm$^{-2}$, and 
$\dot{M} (r, t)$ is the mass accretion 
rate in the unit of g~s$^{-1}$, which is 
defined by $- 2 \pi r \Sigma v_r$, and $v_r$ is the radial 
velocity of the gas flow in the disk, and the source term, 
$s(r)$, is the mass supply rate per unit distance via 
the stream from the secondary star at a given radius of 
the accretion disk, respectively.  
We assume the source term to be time independent 
in the present study, and give our formulation 
for $s(r)$ in the case of tilted disks in Sec.~4.1.

The equation for angular momentum conservation is 
\begin{equation}
\frac{\partial (2 \pi r \Sigma h)}{\partial t} = \frac{\partial (\dot{M} h)}{\partial r} - \frac{\partial}{\partial r}(2 \pi r^2 W) - D + h_{\rm LS}~s, 
\label{ang}
\end{equation}
where $W$ is the vertically integrated viscous stress, 
$D$ is the tidal torque exerted by the secondary star, 
and $h=\sqrt{GM_{1}r}$ and $h_{\rm LS}=\sqrt{GM_{1}r_{\rm LS}}$ 
are the specific angular momentum of the disk matter 
and that of the gas stream from the secondary star 
at a given radius, respectively.   
Here $M_{1}$ is the mass of the primary star, 
and $G$ is the gravitational constant, respectively.
We assume that the specific angular momentum of the gas 
stream is conserved to be that at the Lubow-Shu radius, 
$r_{\rm LS}$ \citep{lub75AD}.  
Here we consider only $r\phi$-component of 
viscous stress tensor, expressed by $w_{r\phi}$, and hence, 
$W \equiv \int - w_{r \phi} dz$.  We take 
$\phi$ as the azimuthal angle in the cylindrical coordinates.  
Here $w_{r\phi}$ includes the shear viscosity 
coefficients, and is expressed as $- 3 \rho \nu \Omega / 2$, 
by using the kinematic viscosity $\nu$.  
By adopting $\alpha$-prescription proposed by 
\citet{sha73BHbinary}, $w_{r\phi}$ is represented as 
$- \alpha P$, where $P$ is the pressure and $\alpha$ is 
the viscosity parameter formulated in Sec.~3.4, 
respectively.  
The vertically integrated viscous stress, $W$, is then 
expressed under the one-zone approximation as follows:   
\begin{equation}
W = \frac{3}{2} \nu \Sigma \Omega = 2 H \alpha P_{\rm c} = \alpha \frac{\mathscr{R}}{\mu_{\rm c}} \Sigma T_{\rm c}, 
\label{w}
\end{equation}
where $T$ is the temperature, 
$\Omega = \sqrt{GM_{1}/r^3}$ is the Keplerian angular velocity, 
$\mathscr{R}$ is the gas constant, 
and $\mu$ is the mean molecular weight, respectively.  
We do not consider the radiation pressure, and $P$ represents 
the gas pressure defined as 
\begin{equation}
P = \frac{\mathscr{R}}{\mu} \rho T. 
\label{pressure}
\end{equation}
Here the scale height of the disk, $H$, is defined as 
\begin{equation}
H = \frac{1}{\Omega} \sqrt{\frac{\mathscr{R}}{\mu_{\rm c}} T_{\rm c}}, 
\label{thickness}
\end{equation}
by using $\frac{P_{\rm c}}{\rho_{\rm c}} = (H\Omega)^2$ on the basis of 
the one-zone approximation.

As for the tidal torque, we adopt the following expression 
by \citet{sma84DI}: 
\begin{equation}
D = c \omega r \Sigma \nu \left(\frac{r}{a} \right)^5, 
\label{tidal}
\end{equation}
where $a$ is the binary separation.  
By substituting equation (\ref{mass}) to equation (\ref{ang}), 
we obtain the following relation: 
\begin{equation}
\dot{M} \frac{\partial h}{\partial r} = \frac{\partial}{\partial r} (2 \pi r^2 W) + D + (h - h_{\rm LS})s.
\label{ang2}
\end{equation}

The equation for energy conservation is 
\begin{eqnarray}
C_{\rm P} \left[ \frac{\partial}{\partial t} (2 \pi r \Sigma T_{\rm c}) - 
\frac{\partial}{\partial r}(\dot{M}T_{\rm c}) - 2 \pi \Sigma \nu_{\rm th} \frac{\partial (r F_r)}{\partial r}  - sT_{\rm c} \right] \\ 
= 2 \pi r (Q^+ - Q^-), \nonumber
\label{ene}
\end{eqnarray}
where $T_{\rm c} (r, t)$ is the temperature at the mid-plane 
of the disk in the unit of K.  
Also, $Q^+$ and $Q^-$ represent the heat generation and 
the radiative loss from the disk surface per unit 
surface area, respectively.  
The thermodynamic quantities of $\mu$ in equation (\ref{w}) 
and the specific heat at constant pressure, $C_{\rm P}$, 
are evaluated at the mid-plane disk.
We adopt the chemical composition of population I stars: 
$X = 0.70$, $Y = 0.27$, and $Z = 0.03$, where $X$, $Y$, 
and $Z$ are the hydrogen content, the helium content, and 
the metallicity, respectively.  
In calculating the thermodynamic quantities,  
we take into account the ionization of hydrogen, and the first 
and second ionizations of helium, and the dissociation of 
the H$_{2}$ molecule in the same way as described in 
\citet{pac69redgiant}.
The second, third, and fourth terms in the left-hand side 
of equation (9) represent those of the advection, 
the thermal diffusion, and the input energy by the gas 
stream from the secondary star, respectively.  
In the term of the thermal diffusion 
$F_{r}$ is the temperature gradient expressed as 
${\partial T_{\rm c}}/{\partial r}$, and $\nu_{\rm th}$ signifies 
the thermal diffusivity.  
Here we consider the dynamical diffusivity only, and 
$\nu_{\rm th}$ is given by $2W/\Omega\Sigma$
\citep{mey84ADtransitionwave,min86DNDI}.  
In the fourth term, we assume the gas stream has 
the same temperature as the pre-existing disk matter 
for simplicity.  
We note here that all variables appearing in these equations 
are now calculated once the two quantities, $\Sigma$ and 
$T_{\rm c}$, at a given $r$ are known.

\subsection{Heating and cooling functions}

To solve the energy conservation denoted by equation 
(9), we need to express $Q^+$ and $Q^-$ as 
the functions of $T_{\rm c}$ and $\Sigma$.  
In this study, we consider the following three terms 
as the heating source $Q^+$:  
\begin{eqnarray}
Q_1^+ &=& \frac{3}{2} W \Omega, \\
Q_2^+ &=& D \frac{\Omega - \omega}{2 \pi r}, \\
Q_3^+ &=& \frac{\beta}{2} \frac{GM_{1}}{r} \frac{s}{2 \pi r}.
\label{heat}
\end{eqnarray}
Here, $Q_1^+$, $Q_2^+$, and $Q_3^+$ represent 
the shear viscous heating, the tidal dissipation, 
and the energy dissipation of the gas stream from 
the secondary star, respectively.  
The fraction of energy dissipation $\beta$ depends 
on the radius, and is calculated in Sec.~4.1.   
The cooling rate is expressed as $Q^- = 2F$.  
The radiative loss function $F$ is obtained by integrating 
the vertical structure of the convective accretion disk 
as calculated in e.g.~\citet{min83DNDI,ham98diskmodel}.  
We, however, use simplified interpolation formulae 
below in this study.  
The thermal equilibrium curve of the disk is composed 
of three branches: a hot branch where hydrogen is fully 
ionized, an intermediate branch where hydrogen is 
partially ionized, and an optically-thin cool branch 
where hydrogen is neutral.  

\begin{figure}[htb]
\begin{center}
\FigureFile(80mm, 50mm){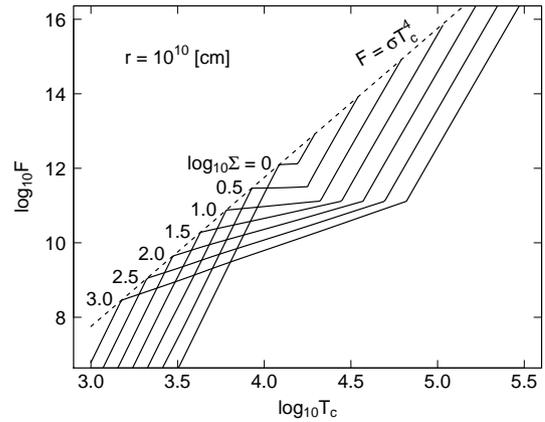}
\end{center}
\caption{The relation between the temperature and the radiative flux as for 6 different values of $\Sigma$ at $r = 10^{10}$~cm, calculated by equations (\ref{fhot}), (\ref{fcool}), and (\ref{fint}).  The dashed line represents $F = \sigma T_{\rm c}^4$.  }
\label{logtlogf}
\end{figure}

In the one-zone model, the radiative flux from the 
optically-thick disk is given as 
\begin{equation}
F = \frac{16\sigma T_{\rm c}^4}{3 \kappa_{\rm c} \rho_{\rm c} H}, 
\label{frad}
\end{equation}
where $\sigma$ is the Stefan-Boltzmann constant, and 
$\tau = \kappa_{\rm c} \rho_{\rm c} H$ is the optical depth of the disk, 
respectively.  
As for the hot branch, the opacity ($\kappa$) is given 
by Kramers law of ionized gas (equation (3.14) in 
\cite{can88outburst}) as follows: 
\begin{equation}
\kappa = 2.8 \times 10^{24} \rho T^{-3.5}~~{\rm cm}^2~{\rm g}^{-1}. 
\label{kramers}
\end{equation}
By combining equations (\ref{thickness}), 
(\ref{frad}), and (\ref{kramers}), the radiative flux 
at the hot branch is obtained as 
\begin{equation}
\log F_{\rm hot} = 8 \log T_{\rm c} - \log \Omega - 2 \log \Sigma - 0.5 \log \mu_{\rm c} - 23.405.
\label{fhot}
\end{equation}
By contrast, we adopt the opacity of the negative hydrogen 
given by the following interpolation formula 
in the optically-thin region on the basis of 
the Cox and Stewart opacity \citep{cox69opacity} as shown 
in Fig.~2 in \citet{can84ADvertical}: 
\begin{equation}
\kappa = 5.13 \times 10^{-19} \rho^{0.62} T^{5.8}~~{\rm cm}^2~{\rm g}^{-1},   
\label{negHopacity}
\end{equation}
where we do not consider the molecular opacity.
The radiative flux in the optically thin case is then as follows: 
\begin{equation}
F = \tau \sigma T_{\rm c}^4.  
\label{fradcool}
\end{equation}
From equations (\ref{thickness}), (\ref{negHopacity}), and 
(\ref{fradcool}), the radiative flux at the cool branch 
is obtained as 
\begin{equation}
\log F_{\rm cool} = 9.49 \log T_{\rm c} + 0.62 \log \Omega + 1.62 \log \Sigma + 0.31 \log \mu_{\rm c} - 25.48.
\label{fcool}
\end{equation}
As for the intermediate branch, we assume that the cool 
branch extends to the critical temperature $T_{\rm A}$ 
at which $F = \sigma T_{\rm A}^4 = F_{\rm A}$, 
and that the hot branch extends to $T_B$ at which 
the radiative flux is approximately represented by 
\begin{equation}
\log F_{\rm B} = 11 + 0.4\log \left(\frac{2.0 \times 10^{10}}{r} \right).
\label{fb}
\end{equation}
Here, $F_B$ is equal to $F_A$ if $F_B < F_A$.  
The radiative flux at the intermediate branch is 
as follows: 
\begin{equation}
\log F_{\rm int} = (\log F_A - \log F_B) \log \frac{T_{\rm c}}{T_B} / \log \frac{T_A}{T_B} + \log F_B.
\label{fint}
\end{equation}
Some examples of the relation between the temperature and 
the radiative flux are given in Figure \ref{logtlogf}.

\subsection{Radial dependence of the viscosity parameter}

\begin{figure}[htb]
\begin{center}
\FigureFile(80mm, 50mm){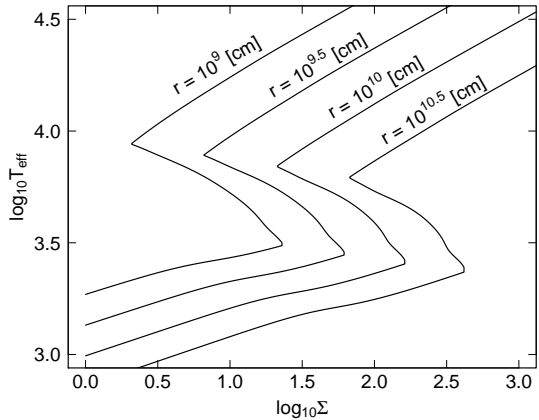}
\end{center}
\caption{The thermal equilibrium curves at $r = 10^{9},~10^{9.5},~10^{10},~{\rm and}~10^{10.5}$~cm, calculated by $2F = Q_1^+ + Q_2^+$.  The vertical axis represents the effective temperature.  Here we use the binary parameters of U Gem, given in Sec.~4.2.  }
\label{scurves}
\end{figure}

The past studies suggest the necessity of $r$ dependence 
of $\alpha$ in equation (\ref{w}) in quiescence and/or 
the difference of $\alpha$ between the hot and cool branches 
in order to reproduce the observational amplitudes of 
outbursts and frequent outside-in outbursts 
\citep{min89quiescenceviscosity,ham98diskmodel,can93DI}.  
We therefore adopt the following formulation: 
\begin{eqnarray}
\log \alpha &=& \frac{1}{2} (\log \alpha_{\rm hot} - \log \alpha_{\rm cool}) \left[1 - \tanh \left(\frac{4-\log T_{\rm c}}{0.4} \right) \right] \\ \nonumber
&+& \log \alpha_{\rm cool}, \\
\alpha_{\rm cool} &=& 0.03 \left(\frac{r}{r_{\rm tidal}} \right)^{0.5}, \\
\alpha_{\rm hot} &=& 0.3, 
\label{alpha}
\end{eqnarray}
where $\alpha$ is the function of $r$ and 
$T_{\rm c}$.  The $\alpha$ value becomes close to 
$\alpha_{\rm cool}$ in the cool state, and becomes close to 
$\alpha_{\rm hot}$ in the hot state.  
We now have the definitions of variables necessary 
for our simulations except for those of $s(r)$ and $\beta$.  
We also obtain the the relation between the surface 
density and the effective temperature, which is called 
the thermal equilibrium curve, as in Figure \ref{scurves}.  
In this figure, we do not consider $Q_3^+$ because $s(r)$ 
depends on the mass input pattern described in Sec.~4.1.  

\subsection{Finite-difference scheme}

The finite-difference scheme used in this paper is 
the same as that described in \citet{ich92diskradius}.  
We treat the conservation of the total 
angular momentum of a disk in the scheme, by letting 
the outer disk edge variable.  
Its detailed description is given in Appendix 1.
We divide the accretion disk into $N$ concentric annuli.  
We define the interface between $i$-th annulus and ($i+1$)-th 
annulus by $r_i$.  
The number of interfaces is then $N+1$ ($i = 0, 1, 2, \dots, N$).  
The inner boundary of the disk is now given by $r_0$ and 
the outer boundary is given by $r_{N}$.  
A special treatment is needed for the outermost annulus and 
its treatment is the same as in \citet{ich92diskradius}. 
In short, the outer edge of the disk, $r_{N}$, is varied in 
such a way to conserve the total mass and the total angular 
momentum of the disk.  The largest radius of the disk is 
the tidal truncation radius, $r_{\rm tidal}$.
When the disk tries to expand beyond the tidal truncation radius, 
we fix the disk radius at the tidal truncation radius by removing 
the extra angular momentum from the disk.  In our calculations, 
200 meshes are used at most.  The details of the scheme and 
the radial distribution of meshes are described in Appendix 1.

\section{Mass input patterns and calculated models}

\subsection{Mass input from the secondary star to a tilted disk}

As described in Sec.~3.1, the key of our simulations is 
to consider how outburst behavior changes when we vary 
the mass supply pattern, $s(r)$, in tilted disks 
where the gas stream enters not only the disk outer edge 
but also the inner disk.  
The location of a bright spot on the disk surface varies 
depending on the position of the secondary star against 
the tilted disk.  
To treat this problem, we have estimated which radius of 
the tilted disk the gas-stream trajectory first collides 
with, while the tilted disk rotates half around 
against the secondary during the half period of negative 
superhumps.
We then obtain the time-averaged mass input pattern 
in the $r$-direction, $s(r)$.  
Finally we have formulated three representative mass input 
patterns in the low tilt case, the moderate tilt case, and 
the high tilt case, respectively.  

\begin{figure}[htb]
\begin{center}
\FigureFile(80mm, 50mm){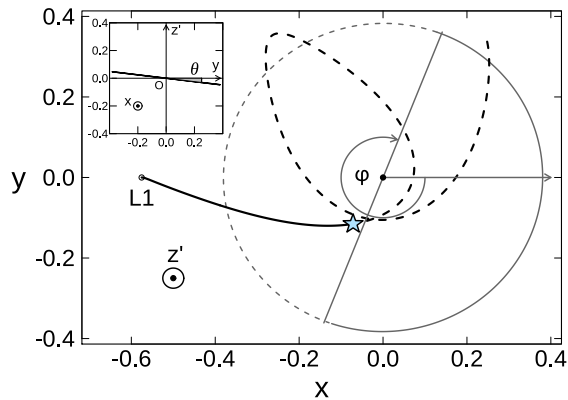}
\end{center}
\caption{Trajectory of the gas stream from the secondary star, which moves on the $x$-$y$ plane.  The grids are normalized by the binary separation.  The L1 point is located at ($x$, $y$) = (0, $-$0.575).  The white dwarf and the secondary are located at ($x$, $y$) = (0, 0) and ($-$1, 0), respectively.  The black point represents the center of the white dwarf.  
Here we adopt the tilt angle, $\theta$ = 7 deg, and we show a particular case of $\varphi$ = 289.6 deg, where $\varphi$ is the angle by which the nodal line (shown by the diametric line) makes with the $x$-axis and it is counted clockwise.   
The solid thick line represents the trajectory, and the mark `star' is the first crossing point of the gas-stream trajectory against the surface of the tilted disk.  The thick dashed line represents the trajectory of gas stream after that, if no collision had occurred.  
The grey thin line represents the contour of the tilted mid-plane disk.  The solid grey line means that the mid-plane is above the $x$-$y$ plane, and the dashed grey line means that it is below the $x$-$y$ plane.  
In the small inlet, we indicate the $y$-$z'$ plane and the tilt angle $\theta = 7$~deg when $\varphi$ = 0.  
}
\label{traj}
\end{figure}

We have computed the ballistic trajectory of a particle 
at first by solving restricted three-body problem 
with the binary parameters of U Gem.  
We have used the equations (1) and (2) in \citet{fla75trajectory}, 
which represent the equation of motion in a co-rotating 
frame with the binary.  
Here we take $x$-$y$ plane as the orbital plane of the binary, 
and $z'$-direction perpendicular to the orbital plane.  
We set the primary and the secondary on $x$-axis as point masses.  
The gas stream comes into the primary Roche lobe via 
the Lagrangian point (the L1 point), and the movement of 
a particle is governed by the gravitational fields of 
the primary and the secondary, and the Coriolis force.  
We assume the initial velocity of a particle toward 
$x$-direction is 0.03, which is normalized by the orbital 
velocity of the binary.  
This value is consistent with the sound speed of the atmosphere 
of the secondary star \citep{lub75AD}.  
The resultant gas-stream trajectory is shown as the thick black 
line in Figure \ref{traj}.  

\begin{figure}[htb]
\begin{center}
\FigureFile(80mm, 50mm){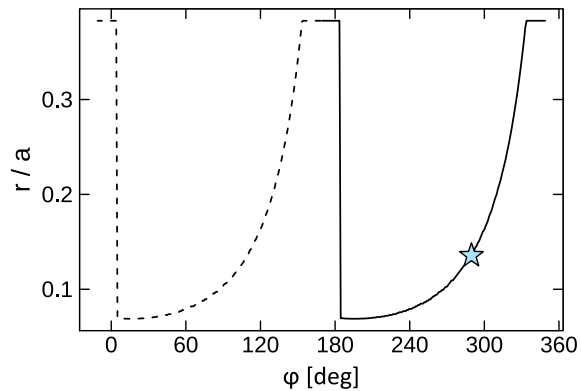}
\end{center}
\caption{
Radial coordinates of the first crossing points of the gas stream from the secondary star when a tilted disk rotates around the $z'$-axis during one period of negative superhumps in the case of $\theta$ = 7~deg.  The horizontal axis represents the rotational angle $\varphi$ of the accretion disk in the co-rotating frame with the binary.  The vertical axis represents the radial distance of the crossing points from the white dwarf, which is normalized by the binary separation.  The mark `star' corresponds to that in Figure \ref{traj}.  The dashed line means the gas stream enters the back face of the tilted disk, and the solid line means it enters the front face, respectively.  
}
\label{traj2}
\end{figure}

We have next determined the geometry of the disk.  
Here we assume that the tilted disk has the thickness 
expected from the steady standard disk for convenience 
\citep{sha73BHbinary} as follows: 
\begin{eqnarray}
\frac{H}{r} = 1.72 \times 10^{-2} \alpha^{-1/10} {\dot{M}_{16}}^{3/20} {\left(\frac{M_{1}}{M_{\solar}}\right)}^{3/8} r_{10}^{1/8} \left[1-\left(\frac{r_{\rm in}}{r} \right)^{1/2} \right]^{3/5},   
\label{height-steady}
\end{eqnarray}
where $r_{10}$ is $r$ in the unit of $10^{10}$~cm, and 
$\dot{M}_{16}$ is $\dot{M}_{\rm tr}$ in the unit of 
$10^{16}$~g~s$^{-1}$, respectively.  
Here $\dot{M}_{\rm tr}$ is the mass transfer rate from 
the secondary star.
We set $\dot{M}_{16}$ and $\alpha$ in equation (\ref{height-steady}) 
to be 10 and 0.3, respectively, and assume that 
the outermost radius of the disk is the tidal truncation 
radius.  
In reality, the thickness and the size of 
the disk vary with time but we do not consider such effects 
here for simplicity.

By using the trajectory and the disk geometry that we have 
prepared, we have calculated the first-crossing points 
between the gas stream and the tilted disk, once we fix 
a tilt angle ($\theta$ = 7 deg in Figure \ref{traj}).  
The tilted accretion disk rotates clockwise around 
the white dwarf (i.e., the $z'$-axis) in the co-rotating 
frame with the binary, 
and it takes one period of negative superhumps for 
one rotation.  The rotational angle is denoted as $\varphi$ 
and it is defined as the angle from the positive $x$-axis 
to the nodal line of the tilted disk and measured clockwise.  
It varies according to $\varphi = \omega_{\rm nSH}~t + 
\varphi_{0}$, where $\omega_{\rm nSH}$ is the angular velocity 
of the tilted disk, and $\varphi_{0}$ is the initial value 
which we choose when the gas stream collides with 
the nodal line of the tilted disk at the disk edge.  
We show the definitions of $\varphi$ and $\theta$ 
in Figure \ref{traj}.  

The gas stream moves on the $x$-$y$ plane.  
For example, the gas stream collides with the tilted disk surface 
at the mark 'star' in Figure \ref{traj} when $\varphi = 289.6$~deg 
and $\theta = 7$~deg.  
When the tilt angle is small, the crossing point is slightly 
deviated from the nodal line since we assume the disk has 
a finite thickness.  
We have calculated and recorded the radial distance of 
the crossing points from the white dwarf during one rotation of 
the tilted disk, by incrementing $\varphi$ by 1 deg from 
$\varphi_{0}$ to $\varphi_{0} + 2\pi$.  
The results for $\theta = 7$~deg are shown in Figure \ref{traj2}.  
Here we show the results as the dashed line if the gas stream 
collides on the back face of the disk, and as the solid line 
if the gas stream collides on the front face where we assume 
that the observer looks at the disk from above, 
for instance, with an inclination angle $i$ = 45 deg.  
Since these two cases (i.e., the back face and the front face) 
are the same with respect to the radial distribution, 
we have calculated the crossing points only when the gas stream 
collides with the disk on the front face during a half period of 
negative superhumps.  

\begin{figure*}[htb]
\begin{center}
\begin{minipage}{0.329\hsize}
\FigureFile(58mm, 50mm){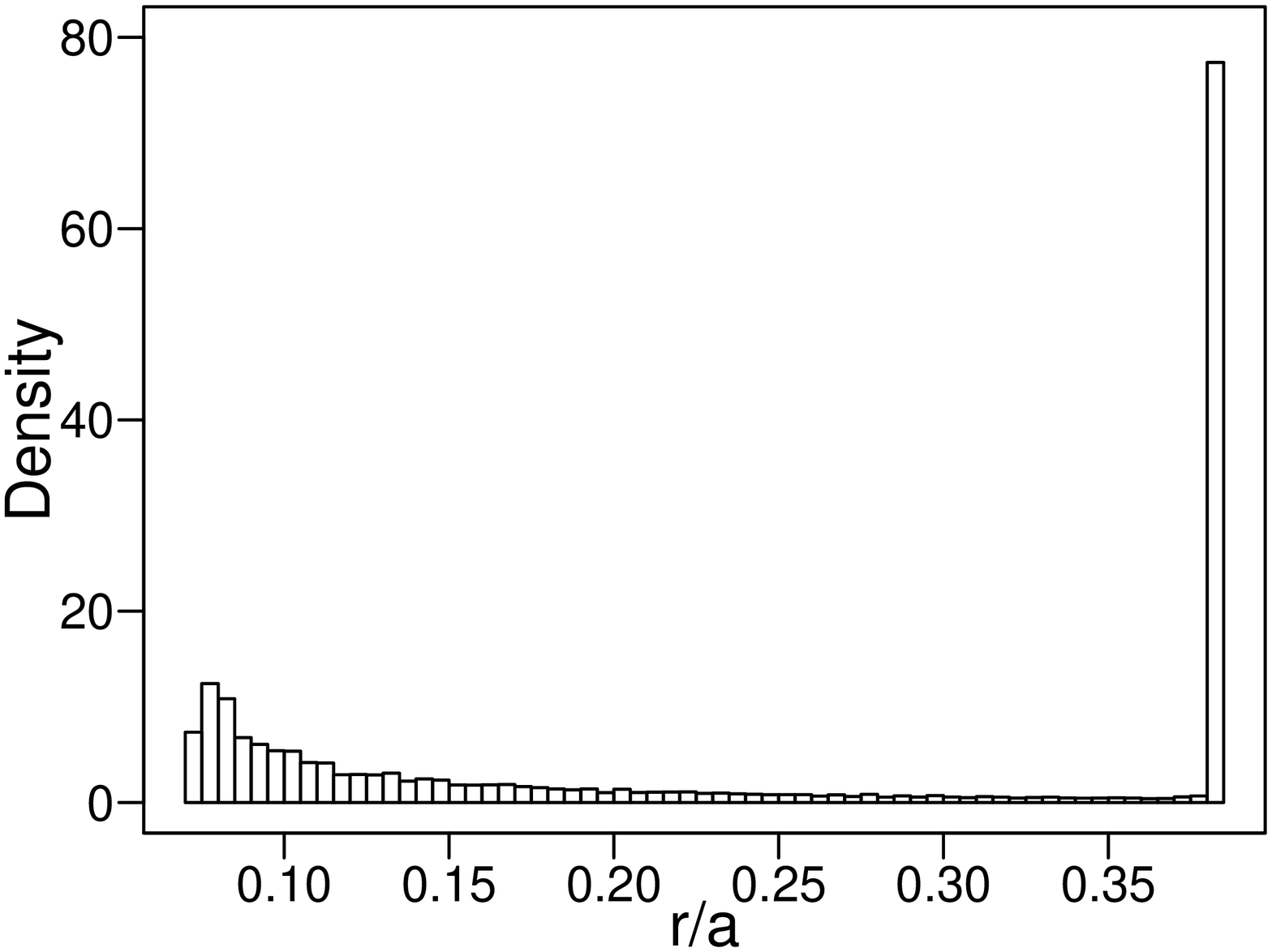}
\end{minipage}
\begin{minipage}{0.329\hsize}
\FigureFile(58mm, 50mm){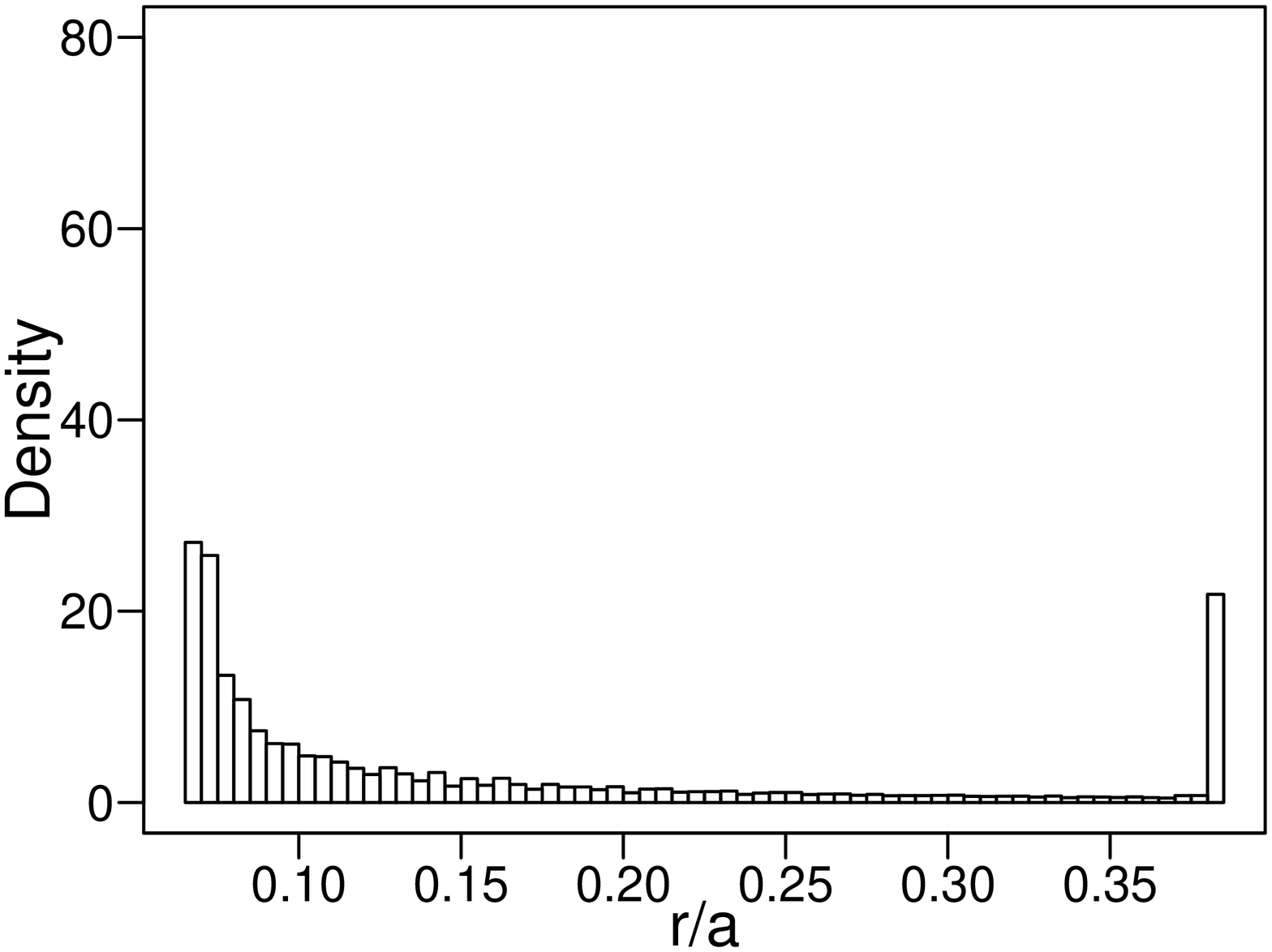}
\end{minipage}
\begin{minipage}{0.329\hsize}
\FigureFile(58mm, 50mm){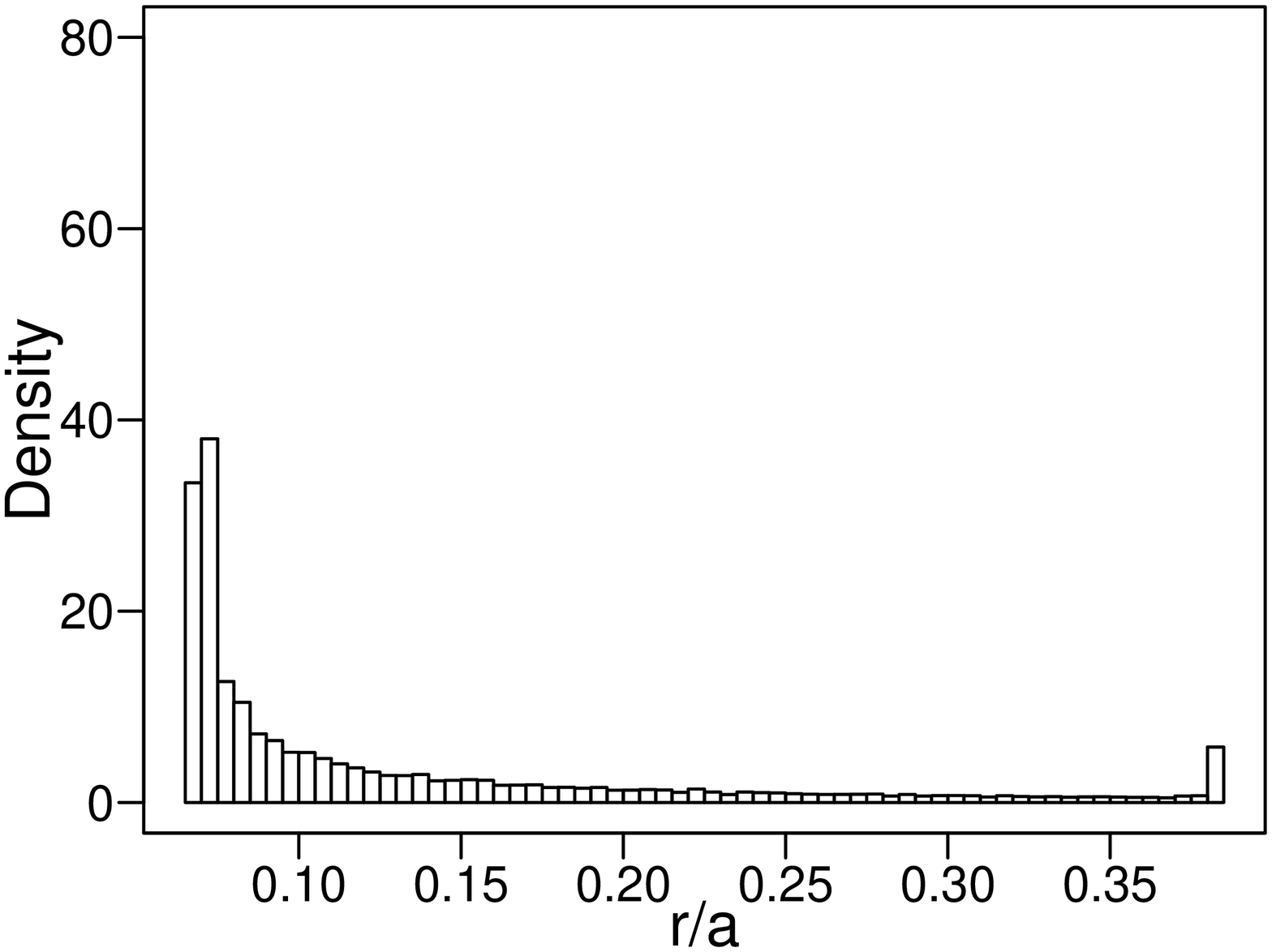}
\end{minipage}
\end{center}
\caption{These panels show how frequent the gas stream enters each annulus during the period of negative superhumps, i.e., while the tilted disk rotates once against the secondary star.  
(Left) The low tilt case with $\theta$ = 3~deg.  (Middle) The moderate tilt case with $\theta$ = 7~deg.  (Right) The high tilt case with $\theta$ = 15~deg.  
}
\label{locations}
\end{figure*}

We have repeated this kind of calculations over the range of 
$\theta$ from 1 to 30 deg by 1 deg.  
We have then estimated how frequent each region of 
the accretion disk in the $r$-direction receives the gas 
stream as shown in Figure \ref{locations}.  
Here we prepare three examples with three different tilt 
angles. These three panels bring the time-averaged 
mass input patterns themselves.  
It has turned out the patterns hardly depend on the tilt 
angle above 15 deg. 
The corresponding tilt angles to these three panels are 
3, 7, and 15 deg in Figure \ref{locations} when we assume 
the standard-disk geometry in the hot disk for convenience.  
However, the actual disk does not always stay in the hot state 
and the disk is mostly thinner than in the case we calculated.  
Therefore these tilt angles do not represent real ones and 
are very much over-estimated.  
In what follows, we refer these three mass input patterns 
as those in the slightly-tilted disk, in the moderately-tilted 
disk, and in the highly-tilted disk, respectively, and we do not 
use these tilt angles any more.  
The mechanism of the disk tilt is still unknown (see also 
Sec.~6.2) and the tilt angle of the accretion disk is 
hard to measure.
There is no negative observational evidence about
high tilt angles.  That is the reason why we consider
the high tilt case as well. 
We see the gas stream often reaches the vicinity of 
$r_{\rm input, min}$ in the high tilt case, while 
it is mostly intercepted at the outer edge in the low tilt case.  
Here $r_{\rm input, min}$ signifies the innermost radius 
where the gas stream from the secondary star reaches.  

\begin{figure*}[htb]
\begin{center}
\begin{minipage}{0.49\hsize}
\FigureFile(70mm, 50mm){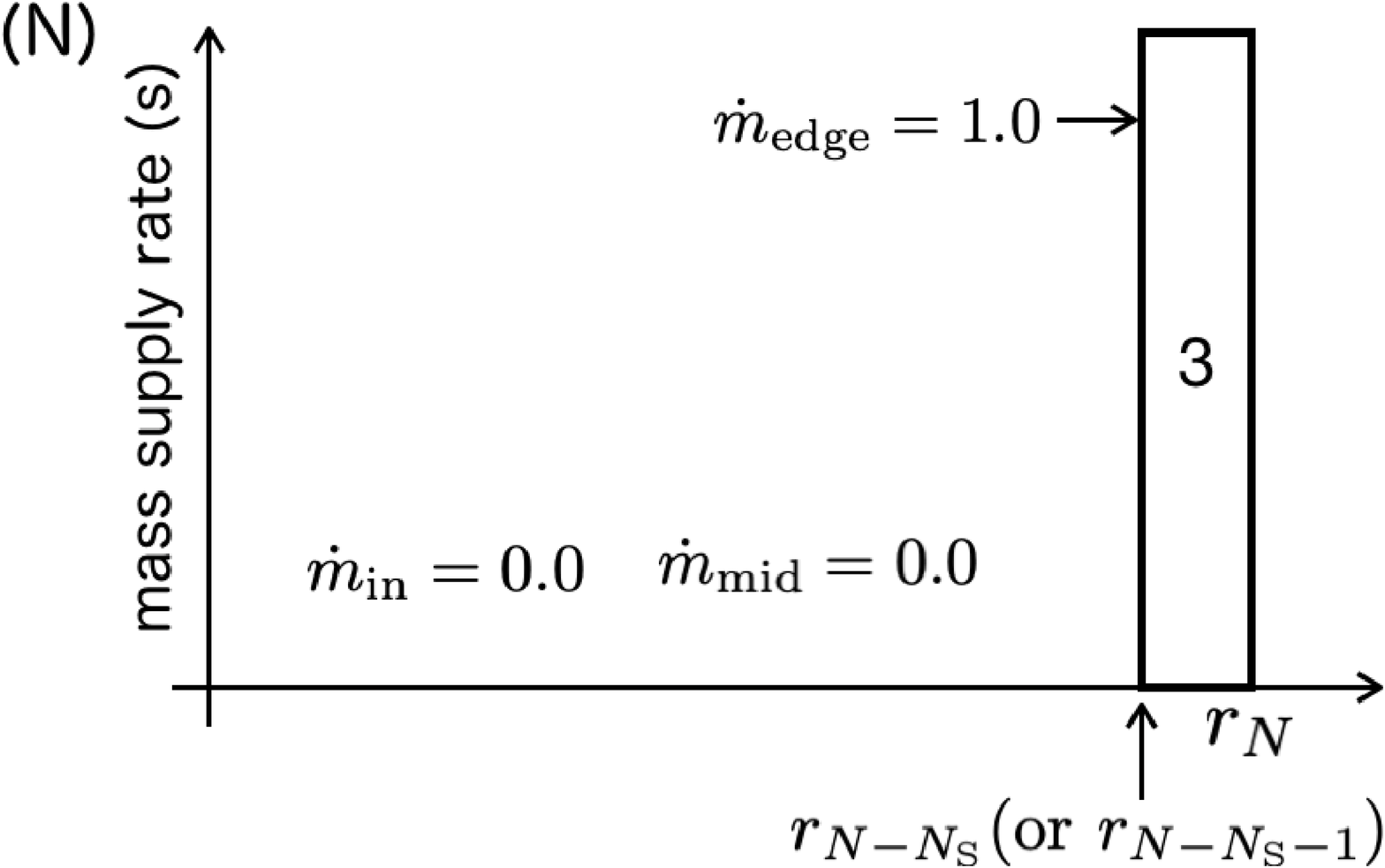}
\end{minipage}
\begin{minipage}{0.49\hsize}
\FigureFile(70mm, 50mm){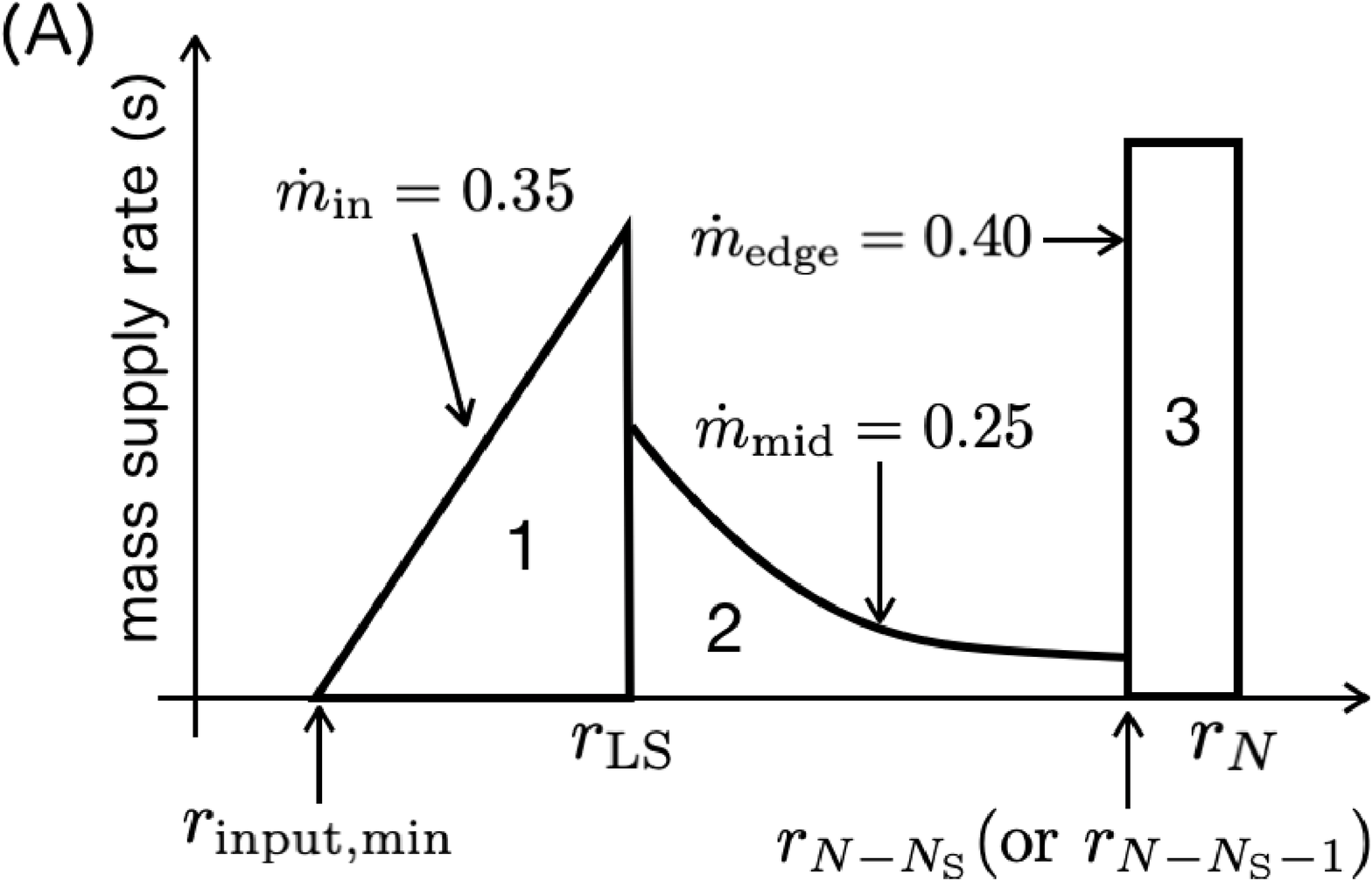}
\end{minipage}
\\
\begin{minipage}{0.49\hsize}
\FigureFile(70mm, 50mm){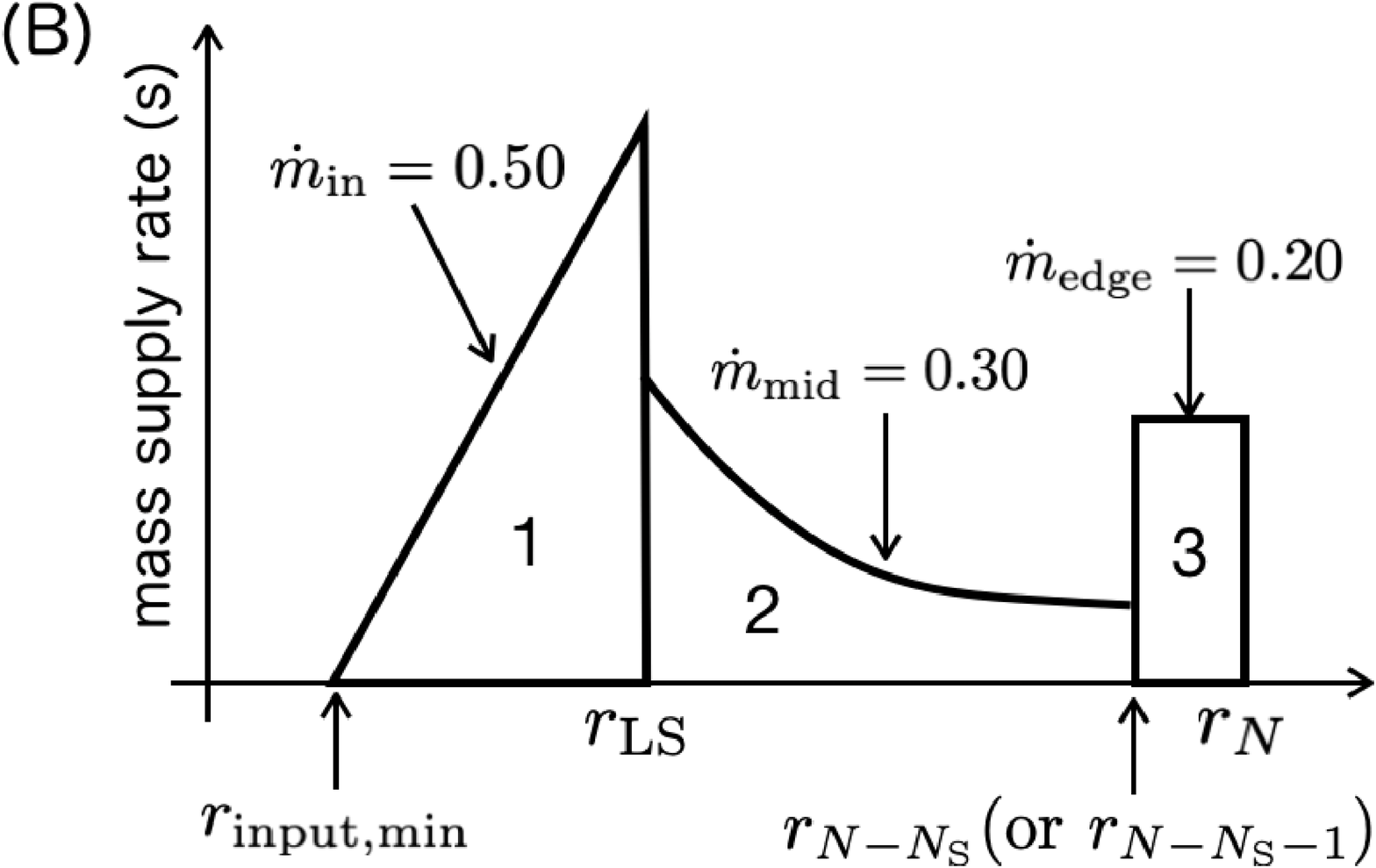}
\end{minipage}
\begin{minipage}{0.49\hsize}
\FigureFile(70mm, 50mm){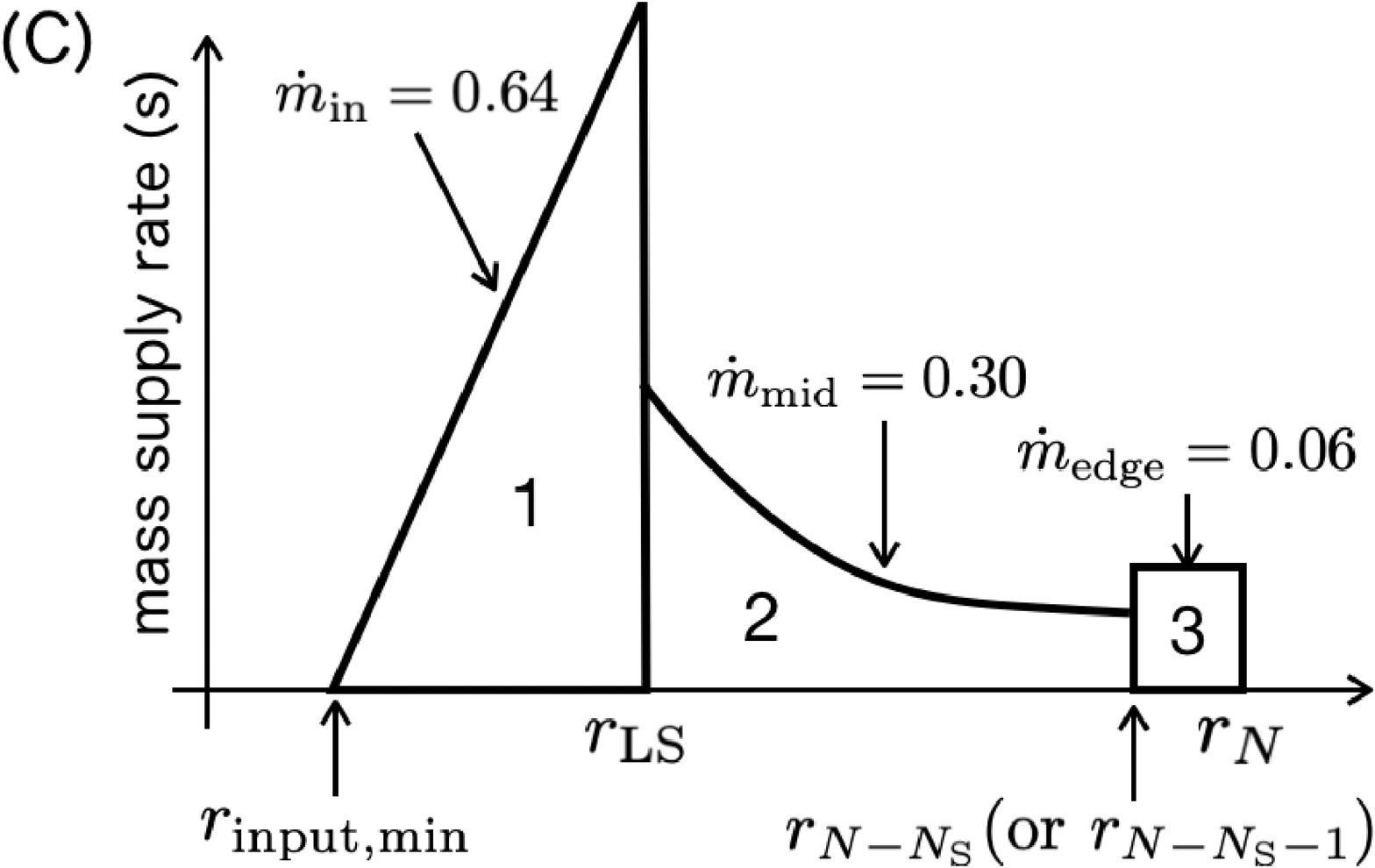}
\end{minipage}
\end{center}
\caption{Schematic pictures of mass input patterns that we use in our simulations on the basis of the results shown in Figure \ref{locations}.  The regions 1, 2, and 3 are the annulus between $r_{\rm input, min}$ and $r_{\rm LS}$, and that between $r_{\rm LS}$ and $r_{N-N_{\rm S}} \textrm{(or}~r_{N-N_{\rm S}-1})$, and that between $r_{N-N_{\rm S}} \textrm{(or}~r_{N-N_{\rm S}-1})$ and $r_{N}$, respectively.  Pattern (N): The non-tilted standard case where the gas stream always enters the outer edge of the disk.  Pattern (A): The low tilt case corresponding to the left panel of Figure \ref{locations}, Pattern (B): The moderate tilt case corresponding to the middle panel of Figure \ref{locations}, Pattern (C): The high tilt case corresponding to the right panel of Figure \ref{locations}.  }
\label{massinput}
\end{figure*}

To implement three mass input patterns shown in Figure 
\ref{locations} to our simulations, we have formulated them 
by dividing the accretion disk into the three regions as in 
each panel of Figure \ref{massinput}.  
We set the three values of $\dot{m}_{\rm in} = 
\dot{M}_{\rm in}/\dot{M}_{\rm tr}$, 
$\dot{m}_{\rm mid} = \dot{M}_{\rm mid}/\dot{M}_{\rm tr}$, and 
$\dot{m}_{\rm edge} = \dot{M}_{\rm edge}/\dot{M}_{\rm tr}$, 
respectively.  
Here, $\dot{M}_{\rm tr}$ is the mass transfer rate 
from the secondary star, and $\dot{M}_{\rm in}$, 
$\dot{M}_{\rm mid}$, and $\dot{M}_{\rm edge}$ are 
the mass input rates into the region 1, 2, and 3, 
respectively.  
We have calculated these values according to the results 
shown in Figure \ref{locations}, and show them in 
Figure \ref{massinput}.  Here we also display the mass input 
pattern of the non-tilted standard case in its upper-left panel.  
In this case, both of $\dot{m}_{\rm in}$ and $\dot{m}_{\rm mid}$ 
are zero.  We thus see the difference in our simulations 
between the non-tilted case and the tilted cases is only 
the mass supply pattern.  
We next have approximately expressed the source term $s(r)$ 
as follows: \\
\begin{eqnarray}
s(r) = \frac{2\dot{M}_{\rm in}}{(r_{\rm LS} - r_{\rm input, min})^2} (r - r_{\rm input, min}) \\ \nonumber
{(r_{\rm input, min} \leq r \leq r_{\rm LS})}, \\
s(r) = \frac{\dot{M}_{\rm mid}}{\log \frac{r_{N - N_{\rm S}}}{r_{\rm LS}}}~\frac{1}{r}~~~{(r_{\rm LS} \leq r \leq r_{N-N_{\rm S}})}, \\
s(r) = \frac{\dot{M}_{\rm edge}}{N_{\rm S}~{\rm dr}}~~~{(r_{N-N_{\rm S}} \leq r \leq r_{N})},
\label{massinputeq1}
\end{eqnarray}
for ${\rm dr}_{\rm N} \geq {\rm dr}$.
Here $N_{\rm S}$ is 10, and ${\rm dr}_{\rm N}$ is the width 
of the outermost radius, respectively.  
\textcolor{black}{If $r < r_{\rm input, min}$, $s(r)$ is 0}.  
When ${\rm dr}_{\rm N}$ is less than ${\rm dr}$ that we define 
in equation (A14) in Appendix 1, the boundary between 
the regions 2 and 3 becomes $r_{N-N_{\rm S}-1}$ instead of 
$r_{N-N_{\rm S}}$ in equations (26) and (27).  
We have chosen the triangular distribution as for the mass 
supply rate at the region 1 in Figure \ref{massinput} 
to avoid numerical difficulties (see discussion in Sec.~6.3).  

\begin{figure}[htb]
\begin{center}
\FigureFile(80mm, 50mm){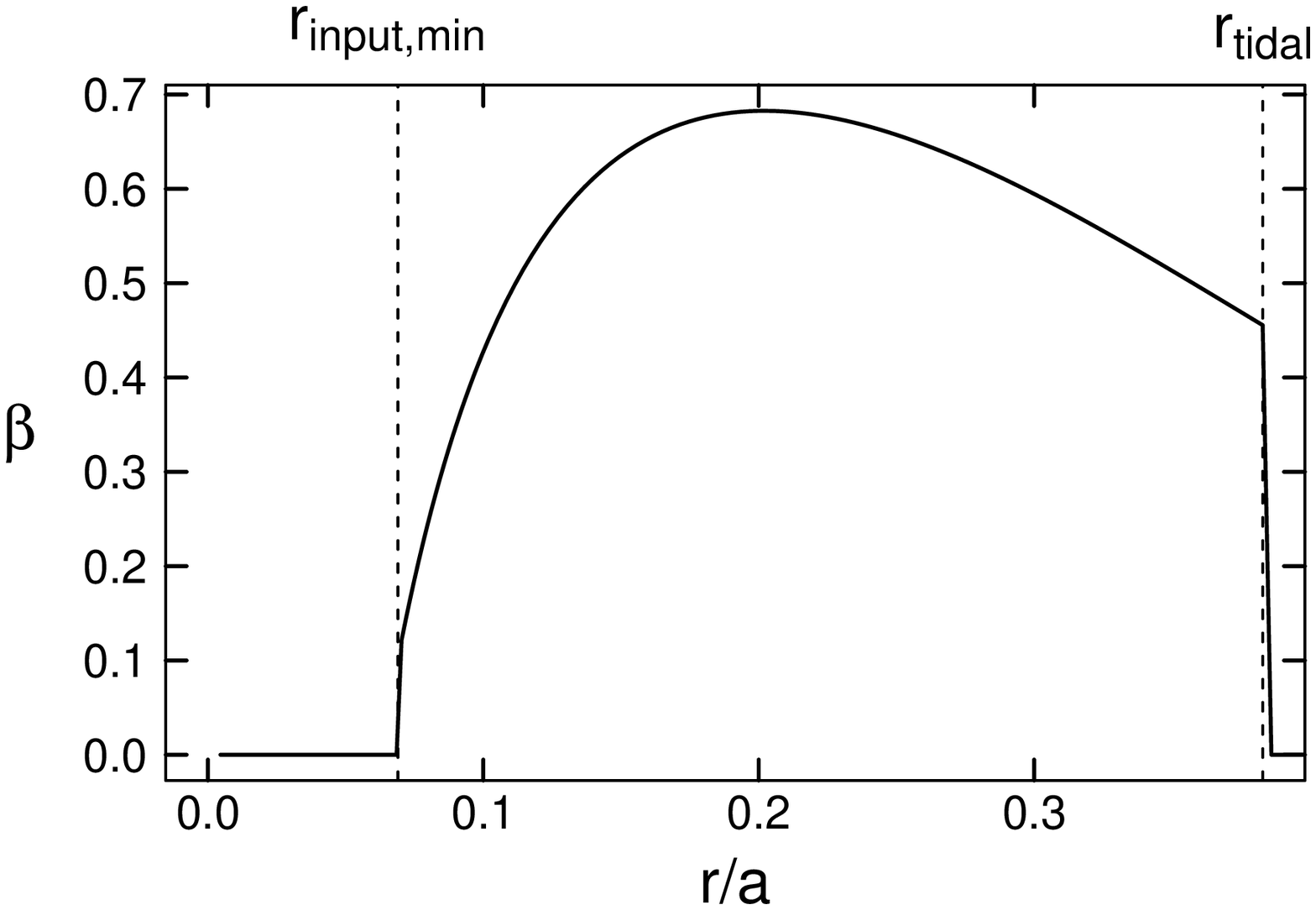}
\end{center}
\caption{Fraction of the thermally dissipated energy by the gas stream with respect to $G M_{1} / 2r$, $\beta$, depending on the radius where the gas stream collides with the tilted disk.  We estimate this value between $r_{\rm input, min}$ and $r_{\rm tidal}$.}
\label{r-beta}
\end{figure}

We have also estimated the value of $\beta$ in equation (12), 
which depends on the radius where the gas stream collides 
with the tilted disk, by calculating the relative speed 
between the gas stream and the disk matter rotating around 
the central white dwarf with the Keplerian velocity.  
If we write the relative velocity as $v_{\rm rel}$, 
the $\beta$ value is defined as follows:
\begin{equation}
\beta = 0.5 \frac{{v_{\rm rel}}^2 / 2}{G M_{1} / 2r}.
\label{beta}
\end{equation}
Here, we assume that the half of the thermally dissipated 
energy by the gas stream is radiated locally as a bright spot, 
and the other half is used to heat the disk.  
The resultant $\beta$ value is given in Figure \ref{r-beta}.

\subsection{Model parameters and lists of calculated models}

We first specify the binary parameters for our simulations.  
Since the aims of the present study is to examine how outburst 
properties depend on the mass input pattern and 
the tilted disk could reproduce the essential features of 
the light variations in IW And-type stars, it is preferable to adopt 
the typical binary parameters in IW And-type DNe 
in our simulations.  Unfortunately, the binary parameters 
of IW And stars are not well known, but 
we have some knowledge that their orbital periods fall on 
a wide range of 0.14--0.48~d above the period gap 
\citep{rod07swsex,arm13aqmenimeri,szk13iwandv513cas,ech07uzser,rin94hxpeg,cie98czaql,gie13j1922,tho04longPDN}.  
We thus can presume that IW And stars would have 
similar binary parameters to those of U Gem stars 
having a similar range of orbital periods except for 
the mass transfer rate.  
Under such a circumstance, 
we basically adopt the binary parameters appropriate for 
U Gem in the following calculations, 
because the period of U Gem is within the period range of 
IW And stars mentioned above and because the binary parameters 
of U Gem are the best known among DNe. 
Besides, we have made additional simulations with the binary 
parameters of KIC 9406652 in order to confirm that our main 
results do not depend very much on the binary parameters used. 
KIC 9406652 is only one object whose mass ratio 
was measured among IW And-type stars \citep{gie13j1922}.  

The binary parameters of U Gem adopted in this study 
is as follows: 
the orbital period ($P_{\rm orb}$) is 0.176906~d, 
the white-dwarf mass ($M_{1}$) is 1.18$M_{\solar}$, 
the mass of the secondary ($M_{2}$) is 0.55$M_{\solar}$, 
the binary separation ($a$) is 1.115$\times$10$^{11}$~cm, 
the tidal truncation radius ($r_{\rm tidal}$) is 0.383$a$, 
the Lubow-Shu radius ($r_{\rm LS}$) is 0.117$a$, and 
the inner edge of the disk ($r_{\rm 0}$) is 5$\times$10$^8$~cm 
(i.e., at the surface of the primary white dwarf), respectively 
\citep{and88DNdiskradius}.  
Additionally, we set $c\omega$ in equation (\ref{tidal}) 
to be 0.4 as a realistic value according to \citet{ich94tidal}.  
According to \citet{lub75AD}, $r_{\rm input, min}$ 
is estimated to be 0.069$a$ when the binary mass ratio ($q$), 
which is defined as the mass ratio of the secondary with respect 
to the primary, is 0.47.  

We next need to specify the mass transfer rate, $\dot{M}_{\rm tr}$. 
Since one of our aims in this paper is to try to understand 
the outbursts of IW And-type DNe, the mass transfer rates, 
$\dot{M}_{\rm tr}$, discussed mainly here are rather high: 
10$^{16.75}$~g~s$^{-1}$ and 10$^{17}$~g~s$^{-1}$ near 
the critical mass transfer rate $\dot{M}_{\rm crit}$, 
above which the disk becomes in the hot steady state.  
However, we also show the results of the cases with other 
two values of mass transfer rates in Appendix 3 
for reference.  

We have shown four mass input patterns including 
the non-tilted, slightly-tilted, moderately-tilted, and 
highly-tilted cases in the previous subsection (see also 
Figure \ref{massinput}), and examine them with a constant 
mass transfer rate.  The models and parameters that we deal with 
are summarized in Table \ref{modeltable}.  
The mass input pattern and the mass transfer rate do not 
change with time within one model.  

\begin{table*}[htb]
\begin{center}
  \begin{tabular}{|c||c|c|c|c|} \hline 
    \textbf{models} & \textbf{mass input pattern}$^{*}$ & \textbf{degree of tilt} & \textbf{$M_{\rm tr}$$^{\dagger}$} & \textbf{figure(s)} \\ \hline \hline
    N1 & N & zero & $10^{16.75}$ & \ref{mdot1675normal}, \ref{mdot1675vband} \\
    A1 & A & low & $10^{16.75}$ & \ref{mdot1675vband} \\
    B1 & B & moderate & $10^{16.75}$ & \ref{mdot1675vband}--\ref{limitcycle3} \\
    C1 & C & high & $10^{16.75}$ & \ref{mdot1675vband} \\ \hline
    N2 & N & zero & $10^{17}$ & \ref{mdot17vband} \\
    A2 & A & low & $10^{17}$ & \ref{mdot17vband} \\
    B2 & B & moderate & $10^{17}$ & \ref{mdot17vband} \\
    C2 & C & high & $10^{17}$ & \ref{mdot17vband} \\ \hline
    N3 & N & zero & $10^{16.5}$ & \ref{mdot165vband} \\
    A3 & A & low & $10^{16.5}$ & \ref{mdot165vband} \\
    B3 & B & moderate & $10^{16.5}$ & \ref{mdot165vband} \\
    C3 & C & high & $10^{16.5}$ & \ref{mdot165vband} \\ \hline
    N4 & N & zero & $10^{17.25}$ & \ref{mdot1725vband} \\
    A4 & A & low & $10^{17.25}$ & \ref{mdot1725vband} \\
    B4 & B & moderate & $10^{17.25}$ & \ref{mdot1725vband} \\
    C4 & C & high & $10^{17.25}$ & \ref{mdot1725vband} \\ \hline
    \multicolumn{4}{l}{$^{*}$See Figure \ref{massinput} for the definition of the mass input patterns.　}\\
    \multicolumn{4}{l}{$^{\dagger}$In the unit of g~s$^{-1}$.}\\
  \end{tabular}
\end{center}
\caption{Summary of the models and parameter sets in our simulations.  }
\label{modeltable}
\end{table*}

\section{Results of numerical simulations}

\subsection{Case of a non-tilted disk (Model N1)}

\begin{figure*}[htb]
\begin{center}
\FigureFile(140mm, 50mm){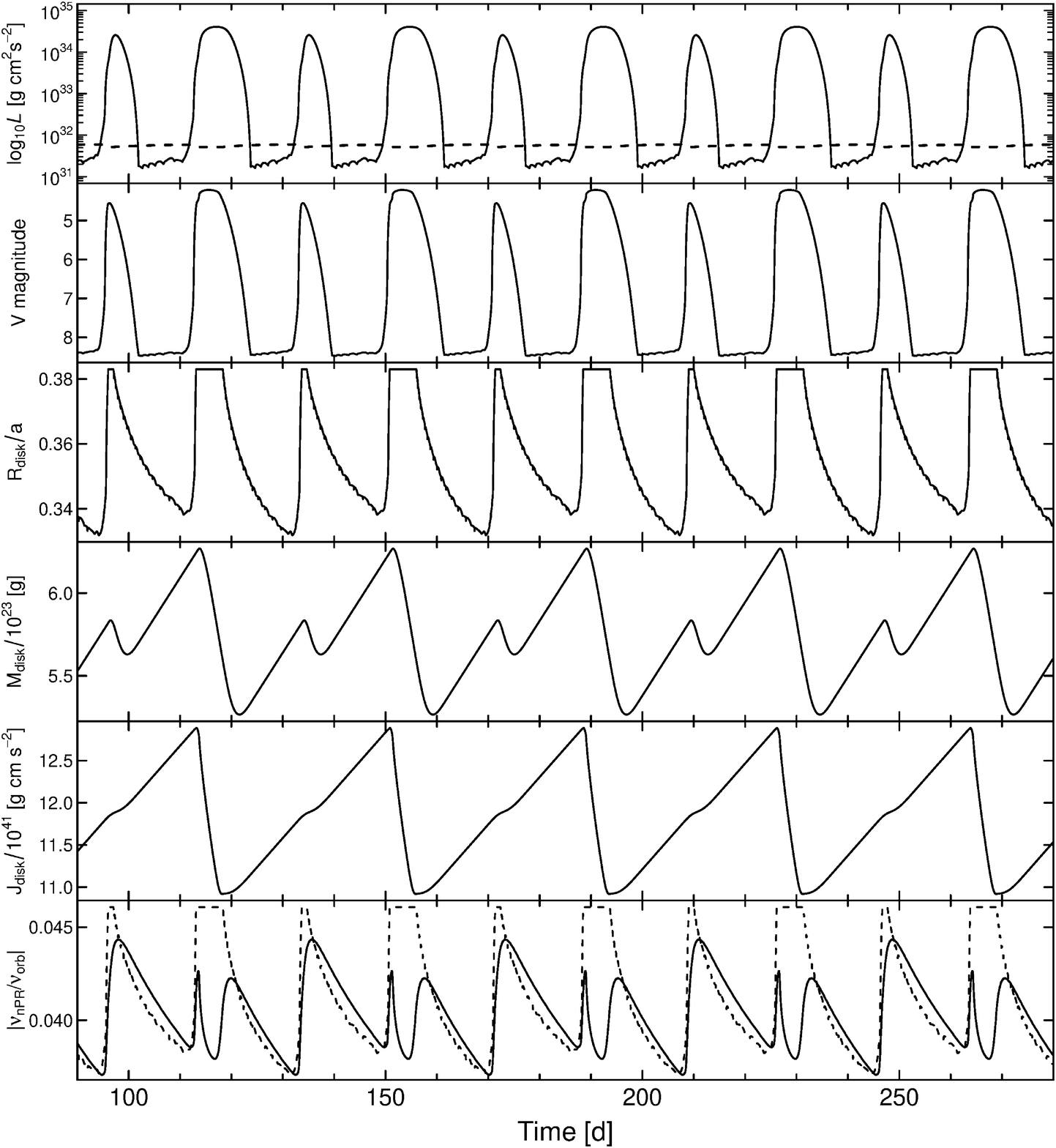}
\end{center}
\caption{Time evolution of the non-tilted accretion disk in the case of $\dot{M}_{\rm tr} = 10^{16.75}$~g~s$^{-1}$ (Model N1 in Table \ref{modeltable}).  From top to bottom: the luminosity of the disk, the absolute $V$-band magnitude, the disk radius in the unit of the binary separation, the total disk mass, the total angular momentum, and the absolute value of the normalized nodal precession rate of the disk.  The dashed line at the top panel represents the luminosity of the bright spot (= $0.25 G M \dot{M}_{\rm tr} / r_{N}$).  The observed luminosity in quiescence is expected not to be lower than this line.  The dashed line in the bottom panel represents the nodal precession rate calculated by equation (\ref{nu_nPR_onlyR}) with $\eta$ = 1.17.  }
\label{mdot1675normal}
\end{figure*}

As mentioned in Sec.~3.1, the non-tilted standard case is 
a special case of our model with the zero tilt angle 
(see also Figure \ref{massinput}).  
It is very convenient to use this case as a test of 
our numerical code. 
We have first calculated this case with $\dot{M}_{\rm tr} = 
10^{16.75}$~g~s$^{-1}$ (i.e., Model N1) by our code. 
The resultant time evolution of the disk is shown 
in Figure \ref{mdot1675normal}.  
In calculating the $V$-band absolute magnitude, we consider 
the disk luminosity and the bright spot, and use 
the same method as that described in \citet{dub18DItest}.  
We set the inclination angle to be 45 deg.  
Here the inclination angle is not the inclination of 
the binary system but the inclination of the disk to 
the observer.
We assume that the disk emission is multi-color blackbody, 
and that the bright spot emits single-temperature blackbody.  
We assume the size of the bright spot is 2\% of the disk, 
and its luminosity is simply approximated to be 
$0.25 G M \dot{M}_{\rm tr} / r_{N}$.  

\textcolor{black}{As seen in Figure \ref{mdot1675normal}, our results 
reproduce outside-in outbursts which are typical of dwarf-nova 
outbursts.  }
The variations of disk luminosity, disk radius, disk mass, 
and disk angular momentum are similar to those in the past 
simulation works 
(e.g., \cite{ich92diskradius,ham00DNirradiation}).  
\textcolor{black}{We see that the resultant light variations regularly 
repeat long and short outbursts.  This is not common in all DNe, 
but is observed in SS Cyg at least in some epochs 
(e.g., during JD 2450100--2450450 in Fig.~1 in \citet{mcg03sscyg}).}
The disk radius suddenly increases at the onset of 
outbursts because of a sudden increase in the angular 
momentum transport from the inner disk to the outer disk, 
while the disk slowly shrinks after the luminosity maximum 
because of addition of mass having low specific angular 
momentum at the outer edge in the cool disk.  
These reproduce the typical observational disk-radius 
variations (e.g., \cite{sma84ugemdiskradius}).  
The disk radius becomes constant around the luminosity 
maxima because of the tidal truncation (see also 
Sec.~3.5 and Appendix 1).  

The bottom panel of Figure \ref{mdot1675normal} represents 
the time variations of the nodal precession rate ($\nu_{\rm nSH}$) 
of a tilted rigid disk.  
If the tilt angle of the disk is very low, the gas stream 
would almost always enter around the disk edge.  Then 
the time evolution of the disk would be almost the same one 
as that in the non-tilted case.  
If we calculate the variations of $\nu_{\rm nSH}$ in this case, 
they are good approximations of those in the very low tilt case.  
If we do not consider the time variation of radial mass 
distribution in the disk, $\nu_{\rm nSH}$ tracks the disk-radius 
variation (the dashed line); however, more realistic variations 
in $\nu_{\rm nSH}$ reflect the time-evolving mass distribution 
(the solid line).  
We give the detailed explanation in Appendix 2.

\subsection{How do the light variations change by the disk tilt ?}

\begin{figure}[htb]
\begin{center}
\FigureFile(80mm, 50mm){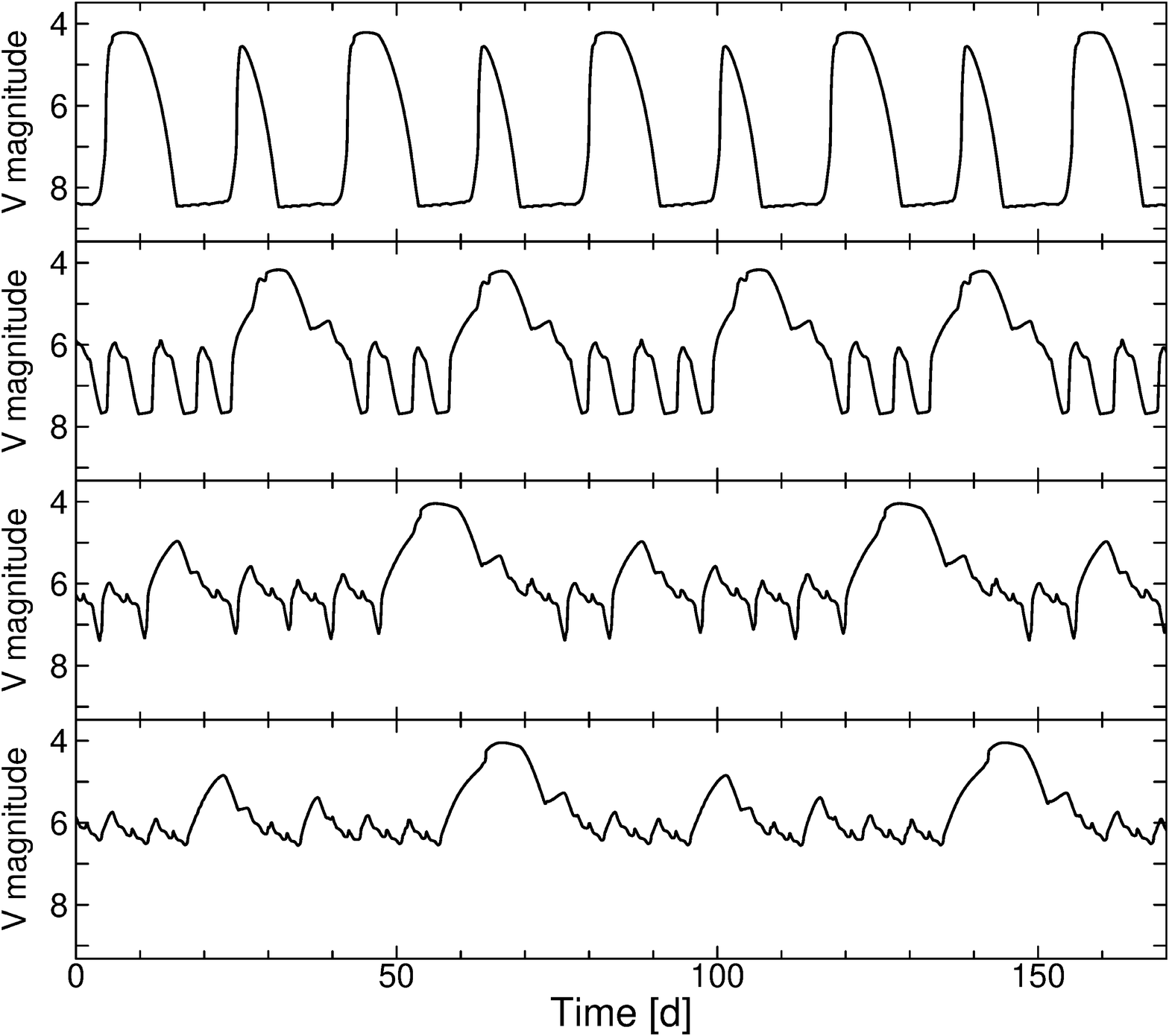}
\end{center}
\caption{Time evolution of the $V$-band magnitude of the tilted accretion disk in the case of $\dot{M}_{\rm tr} = 10^{16.75}$~g~s$^{-1}$.  The contribution of the bright spot is included in our simulations.  From top to bottom: the non-tilted standard case (Model N1), the low tilt case with mass input pattern (A) (Model A1), the moderate tilt case with mass input pattern (B) (Model B1), and the high tilt case with mass input pattern (C) (Model C1).  
}
\label{mdot1675vband}
\end{figure}

Now we are ready to investigate how the light curves change 
from those in the non-tilted standard case when the disk 
is tilted.  
We show the resultant $V$-band light curves of the simulations 
with four mass input patterns in the case of the same 
mass transfer rate as that in the previous subsection 
$\dot{M}_{\rm tr} = 10^{16.75}$~g~s$^{-1}$ (Models N1, A1, B1, and C1) 
in Figure \ref{mdot1675vband}.  
Here we show later parts of simulations in which the effects of 
the initial condition have already faded, and some arbitrary 
offsets of time are made for visibility.  
The luminosity of the bright spot is included in the light 
curves in the tilted cases (Models A1, B1, and C1) 
as in the non-tilted case (Model N1), and it is assumed to be 
$0.25 G M \dot{M}_{\rm tr} / r_{\rm LS}$ for simplicity, 
since the gas stream from the secondary arrives between 
$r_{\rm input, min}$ and $r_{N}$.  
Since the gas stream reaches deeper in the potential well 
in the tilted cases as compared with that in the non-tilted case, 
the level of minima in $V$-band luminosity is thus higher 
in the tilted disk than that in the non-tilted disk 
(see the top and 2nd panels of Figure \ref{mdot1675vband}).  

In Figure \ref{mdot1675vband}, we can see that diverse light 
variations appear in the tilted disk depending on the mass 
input pattern, for the same mass transfer rate.  
As described in the preceding subsection, the disk alternately 
experiences long and short outbursts in the non-tilted disk 
(the top panel).  
In Model A1, the disk alternates between a large outburst and 
a few small outbursts (the 2nd panel).  
In Model B1, large outbursts sandwich mid-brightness interval 
with repetitive dips (the 3rd panel).  
In Model C1, we see repetition of oscillations and brightening 
(the bottom panel).  
The interval between brightening becomes longer and 
the amplitude of outbursts becomes smaller as the tilt becomes 
larger.  
In the tilted cases, we always see slow rise at the onset of 
outbursts (or brightening), which represent inside-out outbursts 
\citep{min85DNDI}.  Since the fraction of gas stream arriving 
in the inner part of the disk becomes larger in the tilted disk, 
the outbursts are easily triggered in the inner disk.  

The light curve in Model B1 shown in the 3rd panel of Figure 
\ref{mdot1675vband} is most interesting, since it shows 
a large outburst and a middle level brightness oscillation 
accompanied with a dip. We therefore examine the results 
in this case in detail.

\subsection{Time evolution of the disk in Model B1}

\begin{figure*}[htb]
\begin{center}
\FigureFile(140mm, 50mm){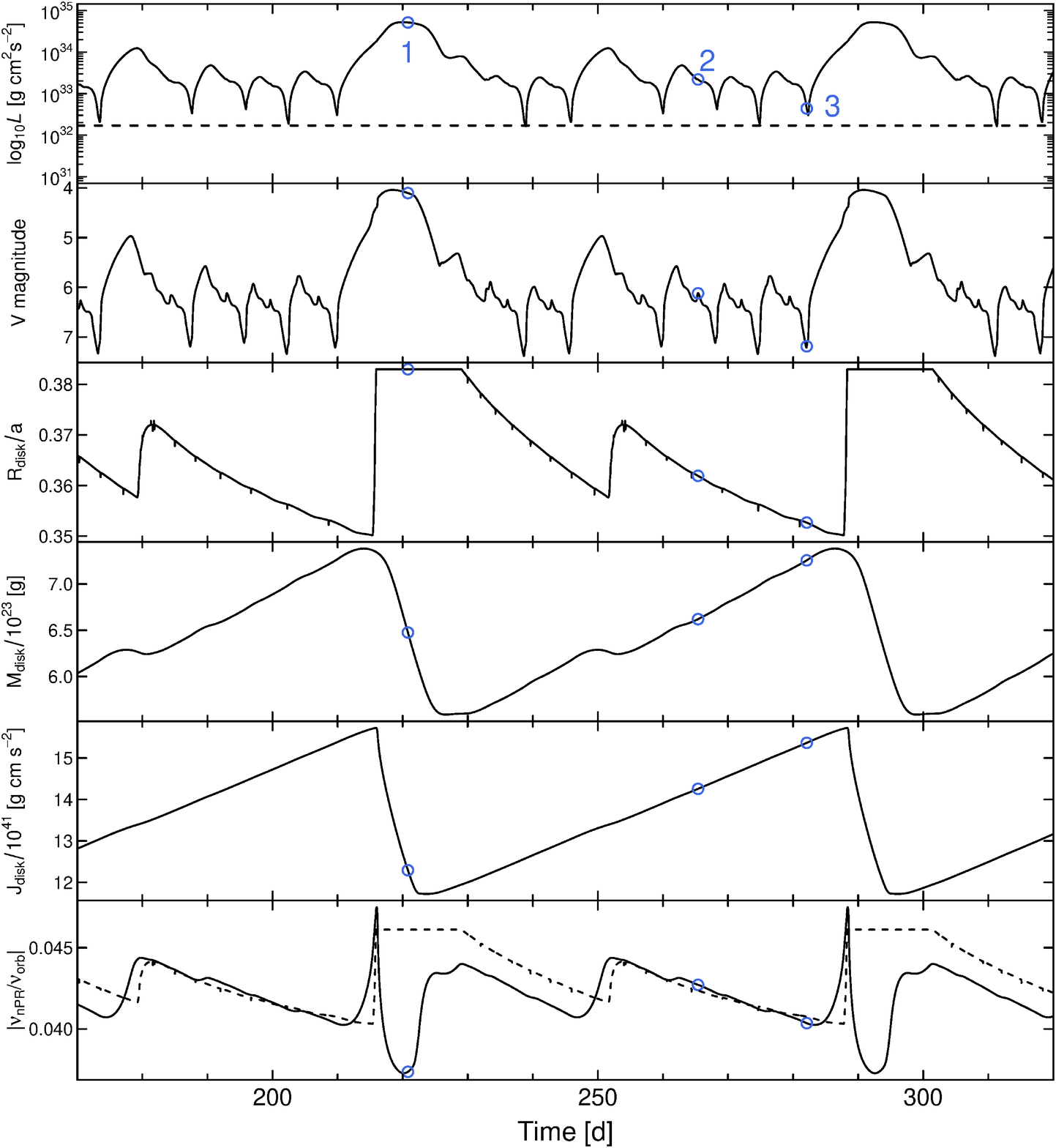}
\end{center}
\caption{Time evolution of the tilted accretion disk in the case of $\dot{M}_{\rm tr} = 10^{16.75}$~g~s$^{-1}$ with mass input pattern (B) displayed in \textcolor{black}{the lower-left panel} of Figure \ref{massinput} (Model B1 in Table \ref{modeltable}).  From top to bottom: the same ones as in Figure \ref{mdot1675normal}.  The circles correspond to the time picked up at each panel in each column of Figure \ref{mdot1675tilt3Tplane}.  We assume that the luminosity of the bright spot is approximately represented as $0.25 G M \dot{M}_{\rm tr} / r_{\rm LS}$.  Also, we use 1.18 as $\eta$ value in calculating the dashed line in the bottom panel.  }
\label{mdot1675tilt3}
\end{figure*}

We show the time-dependent properties of an accretion 
disk in Model B1 in Figure \ref{mdot1675tilt3}.  
The interval during 221--293 days in this figure 
corresponds to one cycle, consisting of a large 
outburst, and oscillatory light variations 
in mid-brightness interval occasionally accompanied 
with dips.  
In the bottom panel of Figure \ref{mdot1675tilt3}, we also 
estimate the variations of the nodal precession rates 
as we do in Model N1.  Their time variations are dominated 
by the time evolution in radial mass distribution 
in the disk (the solid line) rather than the disk-radius 
variations (the dashed line).  
We explain in detail this behavior in Appendix 2.  

\begin{figure*}[htb]
\begin{center}
\begin{minipage}{0.49\hsize}
\FigureFile(80mm, 50mm){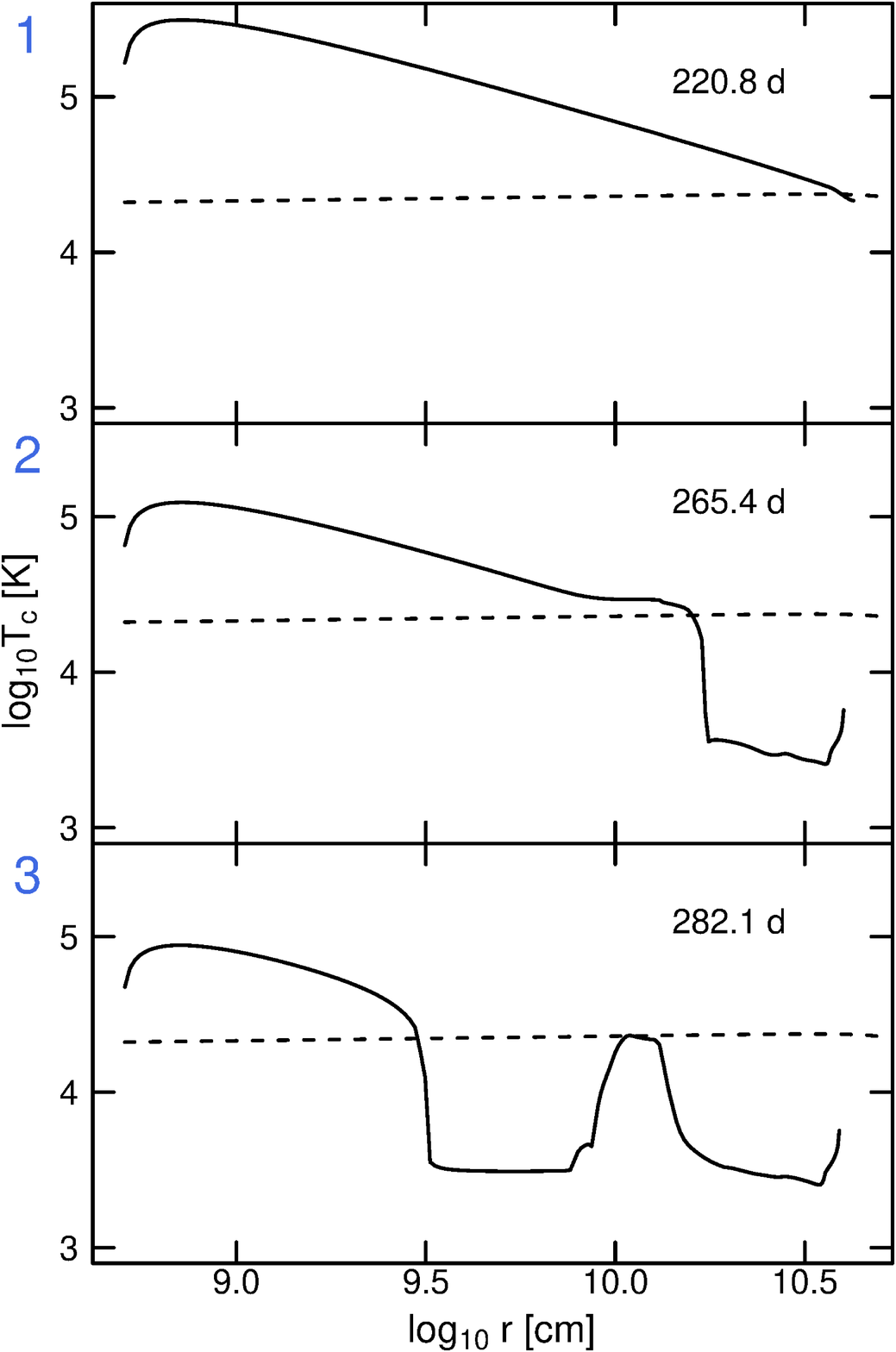}
\end{minipage}
\begin{minipage}{0.49\hsize}
\FigureFile(80mm, 50mm){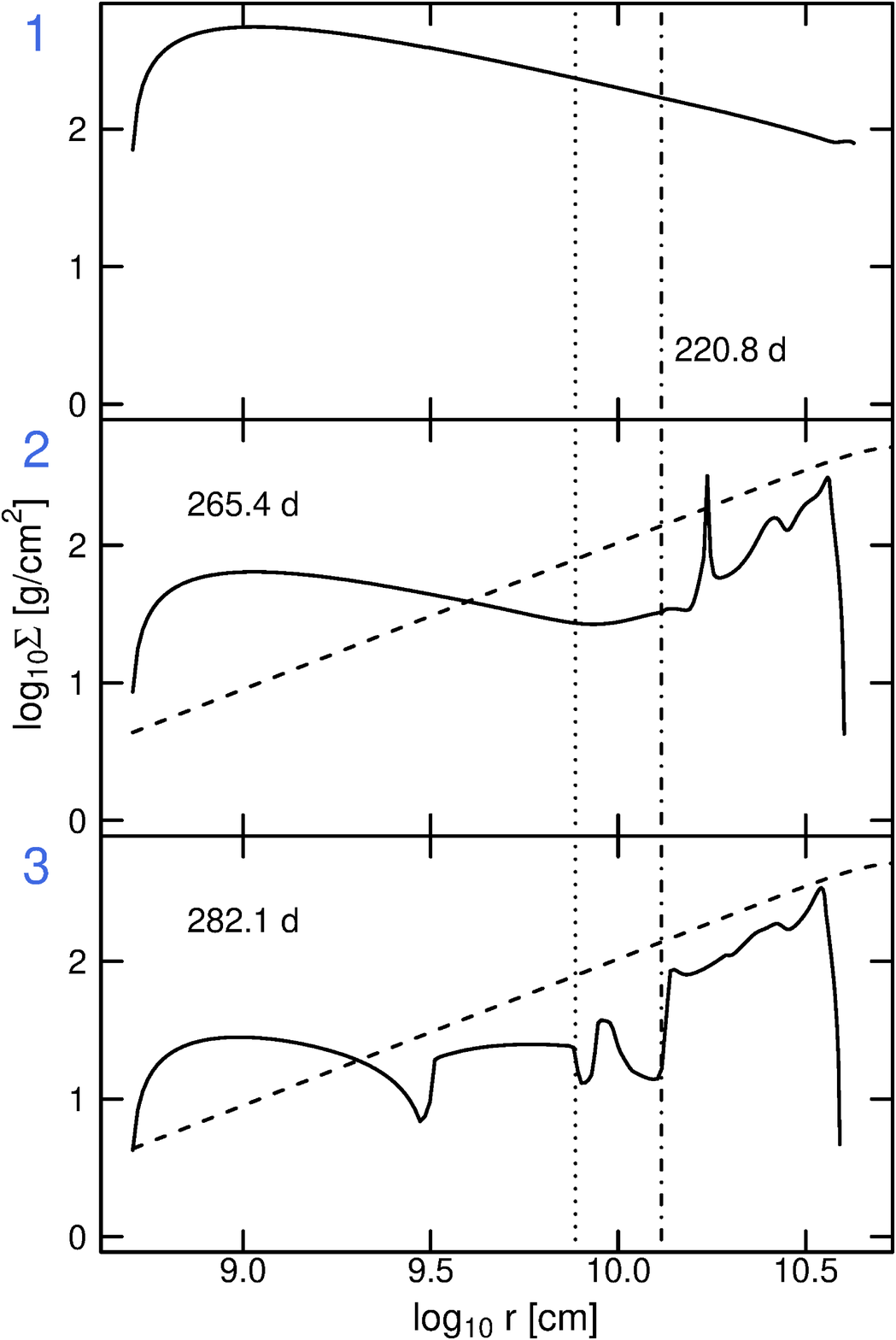}
\end{minipage}
\end{center}
\caption{A part of time evolution of the temperature (left) and the surface density (right) of the disk in the case of Figure \ref{mdot1675tilt3}.  The corresponding time (date) is written in each panel.  They are also marked at the top panel of Figure \ref{mdot1675tilt3}.  The dashed line in the left panel represents the minima of temperature ($T_{\rm hot, min}$) for achieving the hot state as calculated in equation (38) in \citet{ich92diskradius}.  The dashed line in the right panel represents the maxima of surface density ($\Sigma_{\rm cool, max}$) for keeping the disk cool as calculated in equation (35) in \citet{ich92diskradius}.  The dot line and dash-dotted line in the right plane represent $r_{\rm input, min}$ and $r_{\rm LS}$, respectively.  }
\label{mdot1675tilt3Tplane}
\end{figure*}

\begin{figure}[htb]
\begin{center}
\FigureFile(80mm, 50mm){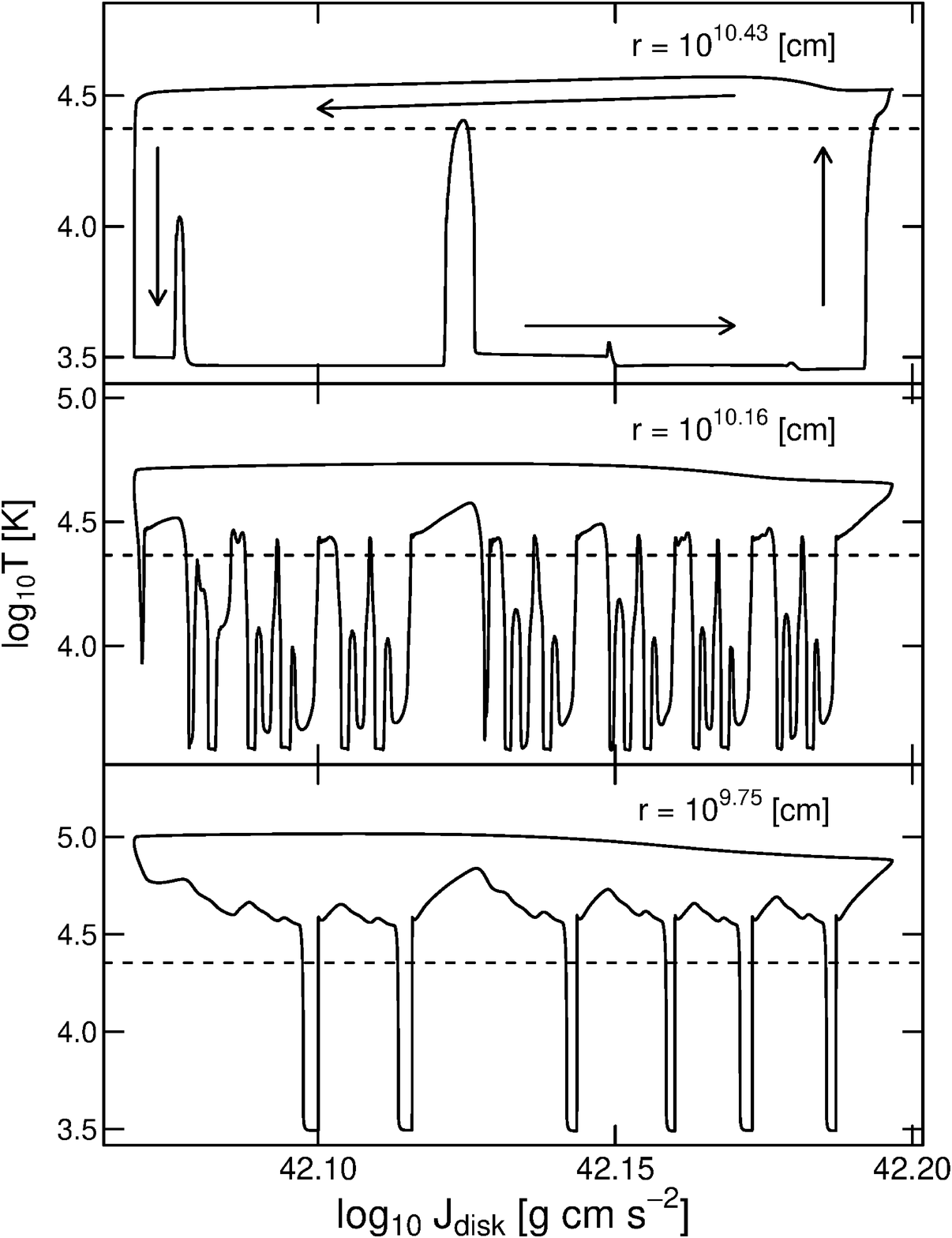}
\end{center}
\caption{$T$-$J_{\rm disk}$ planes at three representative points in the disk, i.e, $\log_{10} r = 10.43, 10.16,~{\rm and}~9.75$, corresponding to the outer, the middle, and the inner parts of the disk, respectively, during 170--320 days.  The arrows in the top panel represent the direction of time evolution.  The dashed line denotes $T_{\rm hot, min}$ defined in Figure \ref{mdot1675tilt3Tplane}.  }
\label{limitcycle3}
\end{figure}

To help understanding this behavior, we show two more figures, 
Figures \ref{mdot1675tilt3Tplane} and \ref{limitcycle3}.  
In Figure \ref{mdot1675tilt3Tplane}, we show three snapshots 
of the distributions of temperature (the left column) and 
surface density (the right column), where times shown 
as three circles in Figure \ref{mdot1675tilt3} (numbered 
as 1, 2, and 3 in its top panel) correspond to those at 
a large outburst, the middle-level brightness interval, 
and a dip, respectively.  
The dashed line in each panel in the left column 
represents the minimum temperature for achieving 
the hot state.  
The dashed line in the 2nd and 3rd panels in the right 
column represents the maximum surface density for 
keeping the entire disk cool.  

Figure \ref{limitcycle3} shows time variations of 
temperature in one cycle at three representative 
radii of the outer, the middle, and the inner parts 
of the disk ($\log_{10} r = 10.43, 10.16,~{\rm and}~9.75$) 
from the top to the bottom, respectively.  
Although the light curve itself is very complicated 
in one cycle, the time variations of the total angular 
momentum in the disk, $J_{\rm disk}$, in the 5th panel of 
Figure \ref{mdot1675tilt3}, show simply slow monotonic 
increase and then rapid monotonic decrease after 
its maximum in one cycle.  We thus use $J_{\rm disk}$ as 
a good indicator of the phase of one cycle. 
The evolutionary tracks for two cycles in Figure 
\ref{limitcycle3} completely overlap and the arrows 
in the top panel represent the direction of time evolution.  

We can understand roughly this behavior by dividing 
the disk into three parts, the inner, the middle, 
and the outer parts.
The inner part spends most of the time in the hot 
state (see, the bottom panel of Figure \ref{limitcycle3}), 
and the outer part stays in the cool state for 
most of the time during one cycle (its top panel), and 
in the middle part between the hot inner part and 
the cool outer part, transition waves (i.e., heating waves 
and cooling waves) frequently go back and forth there 
(its middle panel).  
Thus different regions of the disk behave differently.

At the beginning of one cycle (at brightening), the whole 
disk stays in the hot quasi-steady state (see the 1st panel 
of the left column in Figure \ref{mdot1675tilt3Tplane}).
After the end of a large outburst, the cooling wave 
starting from the outer edge propagates inward and 
it transforms the outer part of the disk into the cool state.
While the outer part gradually accumulates mass in the cool 
state (see, the 4th panel of Figure \ref{mdot1675tilt3}, 
and the 2nd and 3rd panels of the right column in 
Figure \ref{mdot1675tilt3Tplane}), the inner part usually stays 
in the hot state (see the 2nd panel of the left column of 
Figure \ref{mdot1675tilt3Tplane}), and makes 
an intermediate-brightness state.  
Cooling and heating waves alternately develop in the middle 
region (see the middle panel of Figure \ref{limitcycle3}), 
and it causes oscillatory variations in the disk luminosity.  
Occasionally a cooling wave propagating to the inner 
part causes luminosity drop, i.e., 6 dips in the light curve 
correspond to 6 drops in temperature in the inner disk 
per cycle (see the bottom panel of Figure \ref{limitcycle3} 
and the 3rd panels of Figure \ref{mdot1675tilt3Tplane}).  
Finally, a heating wave developed at the middle part 
propagates outward all the way and reaches the outer 
edge of the disk, and it turns the entire disk into the 
hot state (see the top panels of Figure \ref{mdot1675tilt3Tplane}).  
At that time, brightening is regenerated.  
The drain of a large amount of mass onto the central 
white dwarf during the large inside-out outburst returns 
the disk to the starting point of one cycle.  The time 
necessary to accumulate a large amount of mass in 
the outer disk determines the duration of one cycle.  
The disk thus experiences a new type of accretion cycle 
if the gas stream from the secondary penetrates to 
the inner disk.

\subsection{Brief explanations for the light variations in Model A1 and Model C1}

We explore the light curves in the other tilt 
cases for $\dot{M}_{\rm tr} = 10^{16.75}$~g~s$^{-1}$ 
at first on the basis of our investigation in \textcolor{black}{Sec.~5.3}.  
In Model A1, there are clear quiescent states between outbursts 
and the more frequent large outbursts as compared with those in 
Model B1 (see, the 2nd and 3rd panels of Figure 
\ref{mdot1675vband}).  
This is because the inner disk often drops to the cool state 
and keeps cool for longer time due to the smaller mass input to 
the inner disk, and because the outer disk often enters into 
the hot state due to the larger mass input to the outer edge 
of the disk.  
On the other hand, in Model C1, dips seen in Model B1 disappear, 
because the inner disk always stays in the hot state due to 
the sufficient mass supply inside $r_{\rm LS}$ (see the bottom panel 
of Figure \ref{mdot1675vband}).  
The outer disk stays longer in the cool state because 
of reduced mass supply to it, and hence, the duration of 
one cycle becomes longer as compared with that in Model B1.  

\begin{figure}[htb]
\begin{center}
\FigureFile(80mm, 50mm){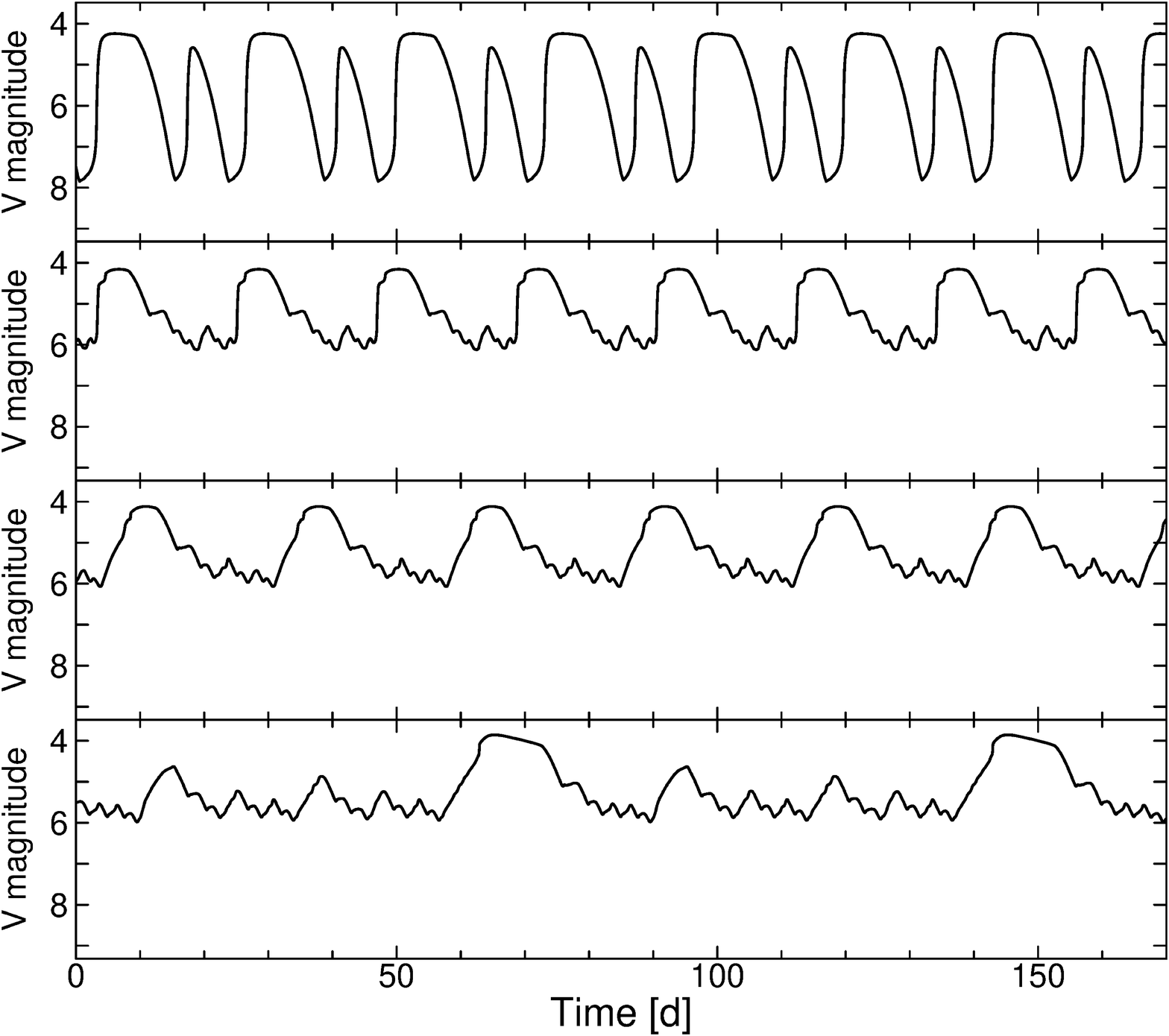}
\end{center}
\caption{Same as Figure \ref{mdot1675vband} but for $\dot{M}_{\rm tr} = 10^{17}$~g~s$^{-1}$ (Models N2, A2, B2, and C2).  }
\label{mdot17vband}
\end{figure}

\subsection{Brief explanations for the light variations in the case of other mass transfer rates}

We next examine the light variations for different mass 
transfer rates.  
The resultant $V$-band light curves for 4 different 
mass input patterns with $\dot{M}_{\rm tr} = 10^{17}$~g~s$^{-1}$ 
(Models N2, A2, B2, and C2) in Figure \ref{mdot17vband}.  
We see the quiescent state completely disappears in 
the tilt cases (Models A2, B2, and C2).  
Under this high mass transfer rate, the mass input rate to 
the inner disk is high enough to keep it hot even in 
Model A2.  
Small-amplitude brightening is repeated in Models A2 and B2, 
since the interval between brightening becomes shorter and 
the mid-brightness interval disappears because of 
the large amount of mass supply to the outer disk edge.  
The behavior in Model C2 is similar to that in Model C1.  
We also explain the other cases in Appendix 3.

\subsection{Test simulations with another set of binary parameters}

\begin{figure}[htb]
\begin{center}
\FigureFile(80mm, 50mm){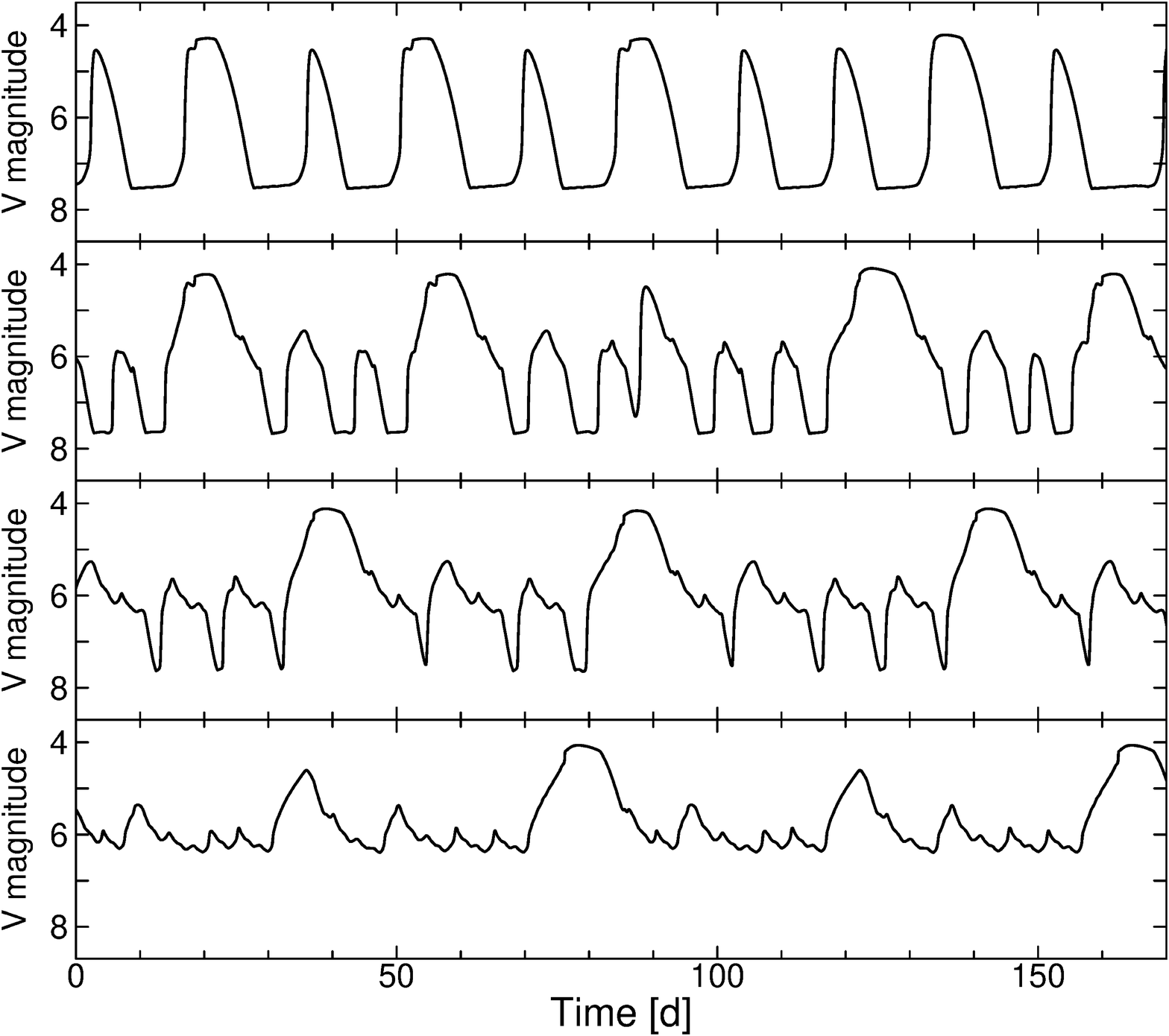}
\end{center}
\caption{Same as Figure \ref{mdot1675vband} but for $\dot{M}_{\rm tr} = 10^{16.9}$~g~s$^{-1}$ and with the binary parameters of KIC 9406652.  }
\label{mdot169vbandkic940}
\end{figure}

We also have checked whether our main results are sensitive for 
the binary parameters or not, by performing some simulations with 
the binary parameters of KIC 9406652 \citep{gie13j1922}.  
The details of the parameter set and the method are 
described in Appendix 4.  
We have tried simulations of the non-tilted standard case 
and the three tilted cases in the case of $\dot{M}_{\rm tr} = 
10^{16.9}$~g~s$^{-1}$ by using the same mass input patterns 
as those shown in Figure \ref{massinput}.  
The resultant $V$-band light curves are exhibited in 
Figure \ref{mdot169vbandkic940}.  
The results are similar to those in Figure \ref{mdot1675vband}, 
and the major features in the simulations with the binary 
parameters of U Gem: frequent outside-in outbursts in 
the non-tilted case, repetition of small inside-out outbursts 
and a large inside-out outburst in the low tilt case, 
repetition of oscillatory state with dips terminated by 
brightening in the moderate tilt case, and that without dips 
in the high tilt case are reproduced.  
We thus have confirmed a new accretion cycle also occurs 
if the system has a tilted disk and a relatively high mass 
transfer rate.

\section{Discussion}

\subsection{Comparison of our simulations with observations}

As described in Sec.~2, there are the two essential 
features of the light variations in IW And-type DNe; 
cyclic light variations with a characteristic pattern, 
which we call the IW And-type phenomenon, and a wide 
variety in the long-term light curves even within one object.  
Our simulations for the tilted disk give the following 
main results with respect to these features of IW And stars.  
\begin{itemize} 
\item  Some of our simulations of the tilted disk can 
produce light curves reminiscent of the IW And-type 
phenomenon: 
regular repetition of the mid-brightness interval with 
oscillations and sometimes dips followed by an outburst 
(brightening) (see, the third and the bottom panels of 
Figure \ref{mdot1675vband}, the bottom panel of 
Figure \ref{mdot17vband}, and the third and the bottom panel 
of Figure \ref{mdot169vbandkic940}).  
This is produced because the inner, the middle, and the outer regions 
behave differently in the tilted disk where the gas stream reaches 
not only the outer disk edge but also the inner disk.  
The inner disk mostly stays in the hot state and it helps to 
sustain the mid-brightness interval.  The outer disk 
spends long time in the cool state, and once it accumulates 
enough mass, an inside-out outburst occurs which terminates 
the mid-brightness interval.   
Cooling and heating waves frequently go back and forth 
in the middle region between the inner hot disk and 
the outer cool disk, which produces oscillatory variations 
in the mid-brightness interval.  
A cooling wave occasionally reaches the inner disk, and a dip 
(a quiescence) then appears in the light curve.  These phenomena 
combined can produce a kind of new accretion cycle in tilted 
disks as suggested in \citet{kat19iwand}.  
\item Our simulations have demonstrated that the thermal 
instability working on the tilted disk can produce a wide 
variety in light curves depending on mass input patterns, 
i.e., the tilt angle and/or the disk geometry, even with 
constant mass transfer rates (see Figures \ref{mdot1675vband}, 
\ref{mdot17vband}, and \ref{mdot169vbandkic940}).  
That is, different mass input patterns make different 
lengths of intervals between brightening and different 
excursion time to the cool state (i.e., the difference 
between clear quiescence, a short dip, and no quiescence 
in the mid-brightness interval).  
For example, the systems can alternate between 
dwarf-nova outbursts, repetition of intermediate-brightness 
state terminated by brightening, which resembles to 
the IW And-type phenomenon, and repetitive brightening 
similar to heartbeat-type oscillations observed in 
HO Pup (see the lower panel of Figure \ref{obsLCs} and 
Figure \ref{mdot17vband}) if the mass transfer rate is 
relatively high and if the mass input pattern varies 
on long timescales.  
\end{itemize}

On the other hand, we have found it difficult to explain 
by our model the details of the observational IW And-type 
light variations.  
For example, luminosity dips sometimes appear just after 
brightening in observations (see also Figures \ref{imeri} and 
\ref{obsLCs}), but in our simulations, dips appear during 
the mid-brightness interval instead of after brightening 
(see the 3rd panel of Figure \ref{mdot1675vband}). 
Also, the amplitudes of brightening are typically less than 
1 mag in observations, but they are $\sim$1-mag larger 
in our simulations.  
Although oscillations in quasi-standstills are sometimes 
moderate in the observational light curves (see also 
Figure \ref{imeri}), our simulations always show large-amplitude 
oscillations in the mid-brightness interval.  
Z Cam-type standstills are sometimes observed in 
long-term light curves as well (see the upper panel of 
Figure \ref{obsLCs}), but our simulations do not reproduce them.  
This is because the inner hot and the outer cool disks coexist 
when the mid-brightness interval is generated by our models.  
The luminosity never settles down in this case, since 
cooling and heating waves are always triggered 
between the hot and the cool regions.  
Since we have performed our numerical simulations 
with a certain set of assumptions, and since we have not explored 
other possible assumptions, it is not expected to reproduce 
all of observational details of IW And-type stars in this work.  
Other possible assumptions to be explored are such as those, 
for instance, more realistic thermal equilibrium curve 
instead of the simple `S'-shaped thermal equilibrium curve, 
other combinations of the viscosity parameters, 
$\alpha_{\rm hot}$ and $\alpha_{\rm cold}$, and 
other forms of mass input patterns, and so on.  
Those attempts are beyond the scope of this paper.

\subsection{Would IW And-type dwarf novae have tilted disks ?}

Our simulations suggest light variations similar to 
the IW And-type phenomenon appear in tilted disks with 
relatively high mass transfer rates which are close to those 
of Z Cam-type DNe (see, the third and the bottom panels of 
Figure \ref{mdot1675vband}, and the bottom panel of 
Figure \ref{mdot17vband}).  
Although the mechanisms for inducing tilted or warped 
structures are controversial, \citet{mon10disktilt} proposed 
that lift may force a disk to be tilted.  
The lift is the same force as that \textcolor{black}{exerted} on the surface area 
of the wings of airplanes.  They showed the tilt by lift happens 
more easily in the objects having higher mass transfer rates.  
Their and our results are consistent with the recent observations 
in that the IW And-type phenomenon is prevalent 
in Z Cam-type DNe (e.g.,~\cite{kat19iwand}).  
Therefore the disk tilt is regarded as a promising assumption 
to explain the IW And-type phenomenon, and we could naturally 
interpret that SS Cyg stars having lower mass transfer rates 
never show this kind of phenomenon.  

As displayed in the bottom panels of Figure \ref{mdot1675normal} 
and \ref{mdot1675tilt3}, we can now calculate the frequency 
variations of negative superhumps (see also Appendix 2).  
The comparison between our calculations and observations 
might give another evidence of the disk tilt and our model.  
Since the time variations in the precession rate reflect 
the changes in mass distribution within the disk, 
we can also discuss the mass distribution in the disk 
through the variations in negative-superhump 
periods with high time-cadence light curves.  
We try these works in our next paper.

\subsection{Possibility of formation of a gap in the disk}

Let us come back to a problem of the mass input pattern 
discussed in Sec.~4.1, particularly about a triangular 
profile in the region 1 (see also Figure \ref{massinput}), 
and we now discuss why we have used such a profile 
in this work. 
Initially we started our simulations by adopting a step 
function in region 1 for the source function, $s(r)$, 
in the same way as in region 3.  However, we had to stop 
our computation because we encountered some numerical 
difficulty around $r_{\rm input, min}$ when the inner 
part of the disk becomes in the cool state.
What happened was that a vacuum (or a gap in the disk) 
seemed to be formed around $r_{\rm input, min}$, although 
our code could not treat such a hole.  

If the inner disk becomes in the cool state, the disk 
matter hardly moves there because of the low viscosity.  
Since the inner part of the disk below $r_{\rm input, min}$ 
has no mass supply from the secondary star, the disk matter 
sits in its place in such a case.  On the other hand, 
the disk matter between $r_{\rm input, min}$ and $r_{\rm LS}$ 
receives mass from the secondary star with the specific 
angular momentum ($\sqrt{GMr_{\rm LS}}$) larger than that of 
the disk matter, and is forced to move rapidly towards 
the Lubow-Shu radius.  
This is because the third term of the right hand side 
in equation (\ref{ang2}) takes a large negative 
value which dominates over the viscous diffusion 
term (i.e., its first term).
In this case, the disk may split into the two parts, 
the inner cool disk and the outer disk having an inner 
boundary around the Lubow-Shu radius.  

To avoid such a difficulty, we have used such a mass 
input profile in the region 1 where the source term, $s(r)$, 
tapers down to zero at $r_{\rm input, min}$. 
In our problem with our interest, the cool state 
in the inner part of the disk is rather short and 
we have been able to avoid that numerical difficulty 
by this treatment.  In fact, even in our treatment 
we can see such a tendency of thinning disk matter 
just above $r_{\rm input, min}$ as seen in the third panel 
of the right column in Figure \ref{mdot1675tilt3Tplane} 
when the inner disk stays in a cool state. 
Although full 3-dimensional hydrodynamic simulations 
may be necessary to confirm this, the formation of a gap 
in the disk seems very likely to occur if the gas stream 
from the secondary penetrates deeply in the cool disk.  
We need to treat such a case in future.

\section{Summary}

We have performed numerical simulations of the disk 
instability of dwarf novae in the case of tilted disks 
by taking into account the various mass supply patterns 
when the gas stream penetrates deeply in the disk 
under some simplifying assumptions.  
We have found that a new kind of accretion cycle can occur 
in tilted accretion disks as suggested in \citet{kat19iwand}.  
This is made possible because the different parts of the disk 
(i.e., the inner, the middle, and the outer parts of the disk) 
can stay in different thermal states in tilted disks.  
The inner disk mostly stays in the hot state while the outer 
disk spends most of time as the cool state but the latter 
eventually makes an outburst when enough mass is accumulated 
there.  
As a result, alternating mid-brightness interval (repeating 
small outbursts) and brightening (a large outburst) are 
reproduced.  The oscillatory light variations in 
the mid-brightness interval are inevitable because cooling 
and heating waves are always triggered in the middle part 
between the hot inner disk and the cool outer disk.  

By this cyclic accretion, light variations reminiscent of 
the IW And-type phenomenon appear when the mass transfer rate 
is relatively high.  The thermal instability 
in the tilted disk would be a plausible model for explaining 
the most characteristic feature of the IW And-type phenomenon.  
Furthermore, we have found that a quite different variety 
in light curve patterns can be produced depending on 
the mass supply pattern in tilted disks even within 
a given mass transfer rate, which could be a key for 
understanding a wide variety in the long-term light curves 
of IW And stars.  
Successful reproduction of the essential features of IW And-type 
light curves is in favor of by the thermal-viscous instability 
in the tilted disk, although we could not reproduce the details 
such as the mid-brightness interval with low-amplitude oscillations 
and Z Cam-type standstills by our model.
This point is left as a future issue.

\section*{Acknowledgements}

This work was financially supported in part 
by the Grant-in-Aid for JSPS Fellows for young researchers (MK) 
and by JSPS Grant-in-Aid for Scientific Research (C) (17K0583, SM).  
We are grateful to Hiromoto Shibahashi.  
We are also grateful to the participants of ``IM Eri campaign''
led by the VSNET Collaboration (Osaka Kyoiku Univ.~team, 
Tonny Vanmunster, Franz-Josef Hambsch, Hiroshi Itoh, 
Crimean Astrophys.~Observatory team, Berto Monard, Kyoto Univ.~team, 
and Shawn Dvorak).  
\textcolor{black}{We would like to express our appreciation to 
anonymous referees.  }

\ifnum0=1
\section*{Supporting information}

Additional supporting information can be found in the online version 
of this article:
Supplementary tables E1, E2, E3, E4 and E5.
\fi

\appendix

\section*{1.~~Details of finite-difference scheme}

As mentioned in the main texts, the finite-difference scheme 
used in this paper is the same as that described 
in \citet{ich92diskradius}.
We divide the accretion disk into $N$ concentric annuli, 
and define the interface between $i$-th annulus and ($i+1$)-th 
annulus as $r_i$, and the center of $i$-th annulus as 
$r_{i-1/2}$ as described in Sec.~3.5.  
We give $\Sigma$ and $T_{\rm c}$ at the center of each annulus 
and these variables are expressed with the half-integer 
subscript $i-1/2$, while the mass accretion rate, 
$\dot{M}$, is defined at the interface of each annulus 
with the integer subscript $i$.
The inner boundary of the disk is now given by $r_0$ 
and the outer boundary is given by $r_{N}$.  In this study 
we choose the inner boundary at the surface of the 
primary white dwarf. 

As for the time integration, we adopt a hybrid method of 
explicit integration for mass conservation and implicit 
integration for the energy equation.  
The mass of the $i$-th annulus, $\Delta M_{i-1/2}$ is then 
given by $\Delta M_{i-1/2}=\pi (r_i^2 - r_{i-1}^2) 
\Sigma_{i-1/2}$ and the equation of mass conservation 
(i.e., equation (\ref{mass})) is now written in 
our finite-difference scheme as 
\begin{eqnarray}
\Delta M_{i-1/2}^{\rm new} = \Delta M_{i-1/2} + (\dot{M}_i - \dot{M}_{i-1})~dt + \dot{M}_{\textrm{s}, i-1/2}~dt,\\
~~(\textrm{for}~i = 1, 2, 3, \dots, N), \nonumber
\label{fd:mass}
\end{eqnarray}
where $dt$ is the time step, and 
$M_{\textrm{s}, i-1/2}=\int_{r_{i-1}}^{r_{i}} s(r) dr$ 
is mass supply rate to the $i$-th annulus.  
The variable with the subscript ``new'' means that 
at the new time step.  
On the other hand, as for the conservation of angular 
momentum, we use equation (\ref{ang2}). 
By integrating equation (\ref{ang2}) over $r_{i-1/2}$ 
and $r_{i+1/2}$, we obtain  
\begin{eqnarray}
\dot{M}_{i} (h_{i+1/2} - h_{i-1/2}) = (2 \pi r^2 W)_{i+1/2} - (2 \pi r^2 W)_{i-1/2} \\
+ [D + (h - h_{\rm LS}) s]_{i} (r_{i+1/2} - r_{i-1/2}) \nonumber \\
(\textrm{for}~i = 0, 1, 2, \dots, N-1), \nonumber
\label{fd:ang2}
\end{eqnarray}
where the last term is defined in equation (10) of 
\citet{ich92diskradius}.  
The finite difference equation for the energy equation 
is written as follows: 
\begin{eqnarray}
C_{{\rm P}, i-1/2}^{\rm new} [\Delta M_{i-1/2}^{\rm new} T_{{\rm c}, i-1/2}^{\rm new} - \Delta M_{i-1/2} T_{{\rm c}, i-1/2}] = \\
C_{{\rm P}, i-1/2}^{\rm new} [(\dot{M} T_{\rm c})_{i} - (\dot{M} T_{\rm c})_{i-1} + \dot{M}_{\textrm{s}, i-1/2} T_{{\rm c}, i-1/2}^{\rm new}]~dt + \nonumber \\
\frac{4 \pi C_{{\rm P}, i-1/2}^{\rm new} W_{i-1/2}^{\rm new}}{\Omega_{i-1/2}} [ r_{i} 
\frac{T_{{\rm c}, i+1/2} - T_{{\rm c}, i-1/2}}{r_{i+1/2} - r_{i-1/2}} - \nonumber \\
r_{i-1} \frac{T_{{\rm c}, i-1/2} - T_{{\rm c}, i-3/2}}{r_{i-1/2} - r_{i-3/2}} ]~dt + \nonumber \\
\pi (r_{i}^2 - r_{i-1}^2) (Q^{+} - Q^{-})_{i-1/2}^{\rm new}~dt \nonumber \\
(\textrm{for}~i = 1, 2, 3, \dots, N), \nonumber
\label{fd:ene}
\end{eqnarray}
where $T_{\rm c}$ on the interfaces of annuli is chosen as 
that at the center of annuli at the upstream side of flow.  

The most important point in this scheme is that the width of 
the outermost annulus is variable, while those of other annuli 
are fixed.  Some special treatment is necessary to handle 
the outermost annulus and in our formulation we vary $r_{N}$ 
at each time step in such a way to conserve the total angular 
momentum of the disk, and its detailed description is given 
in \citet{ich92diskradius} and that paper should be consulted 
(see, their equation (15)).  
The number of annuli $N$ is also variable as a mesh is either 
added or deleted if the width of the outermost annulus, 
$\Delta r_{N}$, exceeds some pre-specified size or shrinks 
below another pre-specified size when the disk expands or 
contracts.  
The way for increasing or decreasing the number of meshes 
is described in Sec.~2.3 of \citet{ich92diskradius}.  

As for the inner boundary condition, we adopt $r_{-1/2} = r_{0}$ 
and $W_{-1/2} = 0$, which means stress free at the inner edge of 
the disk.  
Here the mass accretion rate to the white dwarf, $\dot{M}_0$, is  
approximately calculated from equation (A2) as follows: 
\begin{equation}
\dot{M}_0 = (2 \pi r^2 W)_{1/2} / (h_{1/2} - h_{0}), 
\label{Mdot0}
\end{equation}
where we assume the tidal torque is minute at the innermost 
annulus, and $s_{1/2} = 0$.  

As for the outer boundary condition, we adopt $W_{N} = 0$ and 
$\dot{M}_{N} = 0$, which means no material outside the disk 
and that the mass does not escape from the outer disk edge.  
In this formulation, the outer disk absorbs the angular 
momentum transferred from the inside by its expansion 
when the outer disk enters the hot state, having high 
viscosity.  
When the disk reaches the tidal truncation radius $r_{\rm tidal}$ 
and tries to expand beyond it, we fix the disk radius at that 
radius by removing the extra angular momentum from the disk.  
The extra angular momentum to be removed is calculated 
by using equation (15) in \citet{ich92diskradius}.  

The total mass and total angular momentum of the disk are 
now written as  
\begin{eqnarray}
M_{\rm disk} &=& \int_{r_{0}}^{r_{N}} 2 \pi r \Sigma dr=\Sigma_{i=1}^{N} \Delta M_{i-1/2} , \\
J_{\rm disk} &=& \int_{r_{0}}^{r_{N}} 2 \pi r \Sigma h dr=\Sigma_{i=1}^{N} h_{i-1/2} 
\Delta M_{i-1/2}.
\label{conserve1}
\end{eqnarray}
It has been demonstrated (see, Sec.~2.3 of \citet{ich92diskradius}) 
that the equations of mass and angular momentum conservation 
of the disk are written in our finite-difference scheme as  
\begin{eqnarray}
M_{\rm disk}^{\rm new} - M_{\rm disk} &=& (\dot{M}_{\rm tr} - \dot{M}_{0})~dt \\
J_{\rm disk}^{\rm new} - J_{\rm disk} &=& (h_{\rm LS} \dot{M}_{\rm tr} - h_{0} \dot{M}_{0}
 -D_{\rm total})~dt, 
\label{conserve2}
\end{eqnarray}
where $\dot{M}_{\rm tr}$ is the total mass transfer rate to 
the disk from the secondary star and it is given by  
\begin{equation}
\dot{M}_{\rm tr} = \int_{r_{0}}^{r_{N}} s(r) dr.  
\label{mdot-tr}
\end{equation}
Here, $D_{\rm total}$ is the total tidal torque exerted on 
the accretion disk, and expressed as follows: 
\begin{equation}
D_{\rm total} = \int_{r_{0}}^{r_{N}} D dr.  
\label{Dtotal}
\end{equation}
Equation (A7) means that the variation of the total 
disk mass is determined by the mass supply rate from 
the secondary minus the mass accretion rate onto 
the white dwarf, while the variation in the total angular 
momentum of the disk is determined by the angular momentum 
supply via the gas stream minus the angular momentum loss 
carried with the accreted matter minus the tidal removal of 
angular momentum from the disk (equation (A8)). 
That is, the exact conservation of mass and angular momentum 
in the disk is preserved during the time evolution 
in our numerical scheme.  

The calculations at each time step are proceeded as follows.  
The necessary variables at the old time step are $N$, $r_{N}$, 
$\Sigma_{i-1/2}$, and $T_{{\rm c}, i-1/2}$ ($i = 1, 2, \dots, N$).  
At first, we calculate $\dot{M}_i$ ($i = 0, 1, 2, \dots, N-1$) 
from the variables such as $W$ and $D$ by equation (A2).  
Next we compute ${\Delta M}_{i-1/2}^{\rm new}$ by equation (A1), 
and derive $r_{N}$ at a new time step from equation (15) in 
\citet{ich92diskradius}.  Finally we calculate 
$\Sigma_{i-1/2}^{\rm new}$ by using $r_{N}$ at a new time step, 
and compute $T_{{\rm c}, i-1/2}^{\rm new}$ from equation (A3).  

We determine each time step $\Delta t$ to satisfy the following 
conditions in all of the meshes.  
\begin{eqnarray}
\Delta t \leq \min \left( \frac{0.3 (\Delta r)^2}{\nu},~~\frac{0.3\Delta r (2 \pi r \Sigma)}{|\dot{M}|} 
\right), \\
| T_{\rm c, new} - T_{\rm c} | \leq 0.03 T, \\
| \Sigma_{\rm new} - \Sigma | \leq 0.03 \Sigma.  
\label{timesteps}
\end{eqnarray}
The first one is the Courant condition to secure the numerical 
stability and means that each time step should be smaller 
than the diffusion timescale.  Here, $T_{\rm c, new}$ and 
$\Sigma_{\rm new}$ are the temperature at the mid-plane of 
the disk and the surface density after a new time step, 
respectively.  

In our calculations in this paper, we set the first 90 annulus 
have the same width in logarithmic scales, and that the others 
do in linear scales.\\
For $r \geq 0.1a~(i \geq 111)$, 
\begin{equation}
dr = \frac{r_{\rm tidal} - 0.1a}{90}, 
\end{equation}
and for $r_{\rm input, min} \leq r < 0.1a~(91 \leq i \leq 110)$, 
\begin{equation}
dr = \frac{0.1a - r_{\rm input, min}}{20}, 
\end{equation}
and for $r < r_{\rm input, min}~(i \leq 90)$, 
\begin{equation}
\log(r_{i}) - \log(r_{i-1}) = \frac{\log(r_{\rm input, min}) - \log(r_{\rm 0})}{90}.
\end{equation}
The reason why we divide the region where 
$r_{\rm input, min} \leq r < 0.1a$ into finer meshes 
than those in the nearby region is to smooth mass 
input in this region (see also Sec.~6.3).

\section*{2.~~Frequency variations of the nodal precession rate}

We have calculated the variations of the nodal precession 
rate of a tilted rigid disk, and show them in the bottom 
panels of Figures \ref{mdot1675normal} and \ref{mdot1675tilt3}.  
The precession rate normalized by the orbital frequency 
is estimated by the following equation 
\citep{pap95tilteddisk,lar98XBprecession}: 
\begin{equation}
|\epsilon^{*}_{-}| = |\nu_{\rm nPR}/\nu_{\rm orb}| = \frac{3}{4} \frac{G M_{2}}{a^3} \frac{\int \Sigma r^3 dr}{\int \Sigma \Omega r^3 dr} \cos \theta, 
\label{nu_nSH}
\end{equation}
where $\nu_{\rm nPR}$ is the nodal precession rate of 
a tilted disk and $\nu_{\rm orb}$ is the orbital frequency, 
respectively.  
Here, we set $\theta$ to be 0.  Also, $\epsilon^{*}_{-}$ 
is always negative, which means the retrograde precession.  
We also calculate $|\epsilon^{*}_{-}|$ depending only 
on the disk radius by assuming $\Sigma (r) \propto r^{-3/4}$, 
which comes from the standard disk \citep{sha73BHbinary},  
via the following equation introduced in 
\citet{osa13v344lyrv1504cyg}: 
\begin{equation}
|\epsilon^{*}_{-}| = |\nu_{\rm nPR}/\nu_{\rm orb}| = \eta \frac{3}{7} \frac{q}{\sqrt{1+q}} \left( r_{N}/a \right)^{3/2} \cos \theta, 
\label{nu_nPR_onlyR}
\end{equation}
where $\eta$ is a correction factor, which typically ranges 
from 0.8 to 1.2 (see Appendix 1 in \cite{osa13v344lyrv1504cyg} 
for details).  
The estimated values by equation (\ref{nu_nSH}) are displayed 
as solid lines and those by equation (\ref{nu_nPR_onlyR}) are 
given as dashed lines in the bottom panels of Figures 
\ref{mdot1675normal} and \ref{mdot1675tilt3}.  

We first explain the variations in Model N1 (see the bottom 
panel of Figure \ref{mdot1675normal}).  
We see the precession rate estimated by equation (\ref{nu_nSH}) 
(the solid line) well traces the expansion/contraction of 
the disk radius (the dashed line) in the short outbursts, 
but that it does not in the long outbursts, 
because $\Sigma (r)$ is drastically altered.  
At the onset of long outbursts, the whole disk enters 
the hot state and a large amount of mass accumulated at 
the outer disk is transported to the inner disk soon after 
the disk-radius expansion.  
The disk stays in the hot state for a while, and finally 
the cooling wave develops from the outer edge.  
It propagates inwards, and simultaneously sweeps up 
disk matter outward across the cooling front.  
Thus the complex changes in the nodal precession rate, i.e., 
the consecutive sets of the rapid increase and decrease and 
then again increase, reflect the three phenomena superimposed: 
the disk-radius expansion, the inward transport of a large 
amount of mass, and the redistribution of the disk mass 
by the cooling wave.  

We next explain the variations of the nodal precession 
in Model B1 (see the bottom panel of Figure \ref{mdot1675tilt3}).  
They show cyclic variations, and large and rapid 
variations occur around the large outburst.  
This kind of violent variations are very similar to 
those in the long outbursts in Model N1, but 
there exists one difference between these two cases, 
that is, the delay of the expansion of the disk to 
the increase of the nodal precession rate at the onset 
of the large outburst (see also the bottom panel of 
Figure \ref{mdot1675normal}).  
This is because the outward propagation of 
the heating wave precedes the disk-radius expansion 
in inside-out outbursts, 
i.e., the mass redistribution by the heating wave occurs 
before the disk expansion, while in long outside-in 
outbursts in the case of non-tilted disks, the heating-wave 
propagation from the outer edge occurs almost simultaneously 
with the disk-radius expansion.
Interestingly, the changes in the surface-density 
distribution near the outer edge of the disk govern 
the nodal precession rate rather than the disk-radius 
variations.  
On the other hand, the precession rate decreases during 
small outbursts in parallel with the disk-radius variations, 
unless the heating wave reaches the outermost region.

\section*{3.~~Brief explanations of the results in models of the lowest and the highest test mass transfer rates}

\begin{figure}[htb]
\begin{center}
\FigureFile(80mm, 50mm){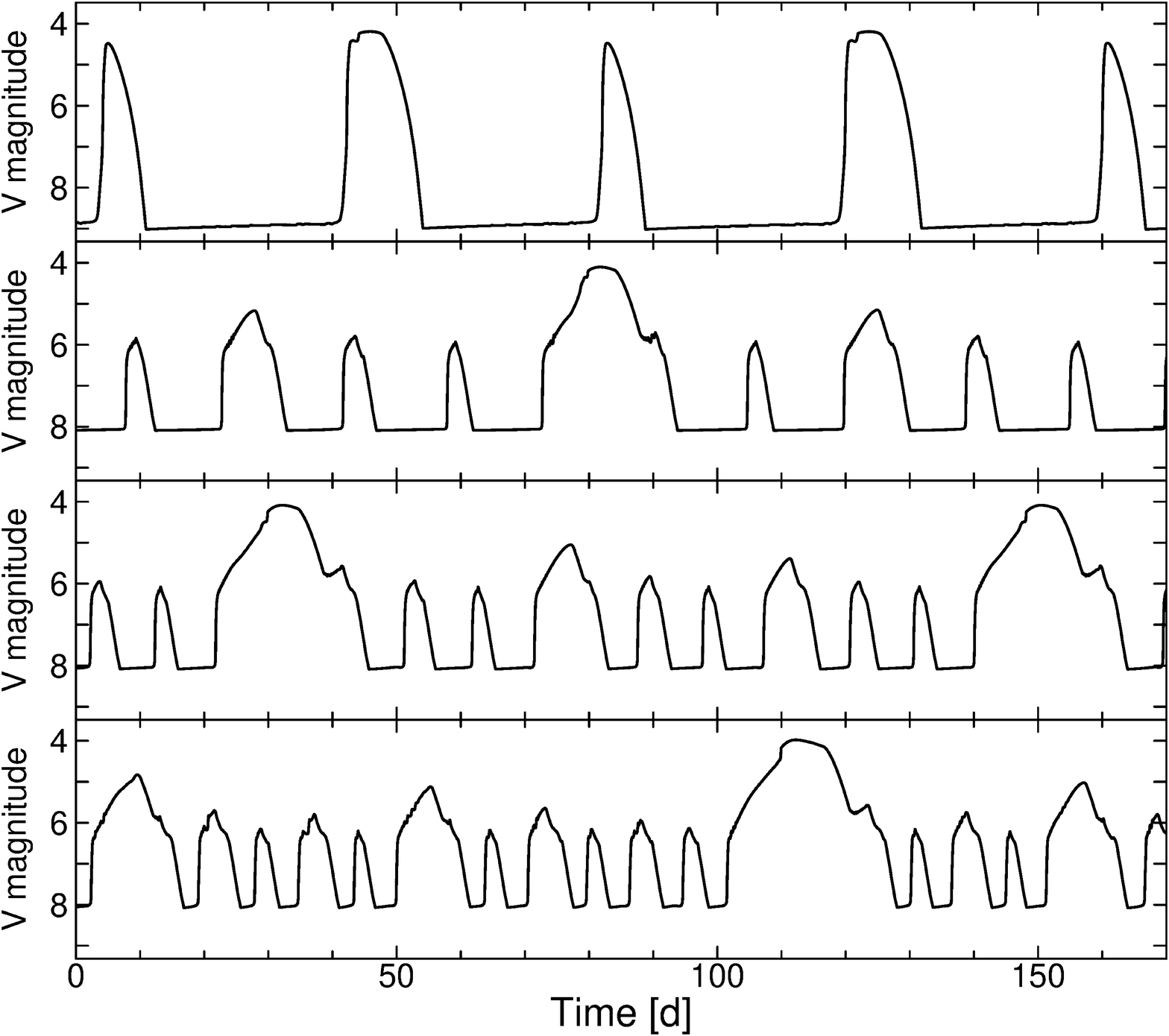}
\end{center}
\caption{Same as Figure \ref{mdot1675vband} but for $\dot{M}_{\rm tr} = 10^{16.5}$~g~s$^{-1}$ (Models N3, A3, B3, and C3).  }
\label{mdot165vband}
\end{figure}

\begin{figure}[htb]
\begin{center}
\FigureFile(80mm, 50mm){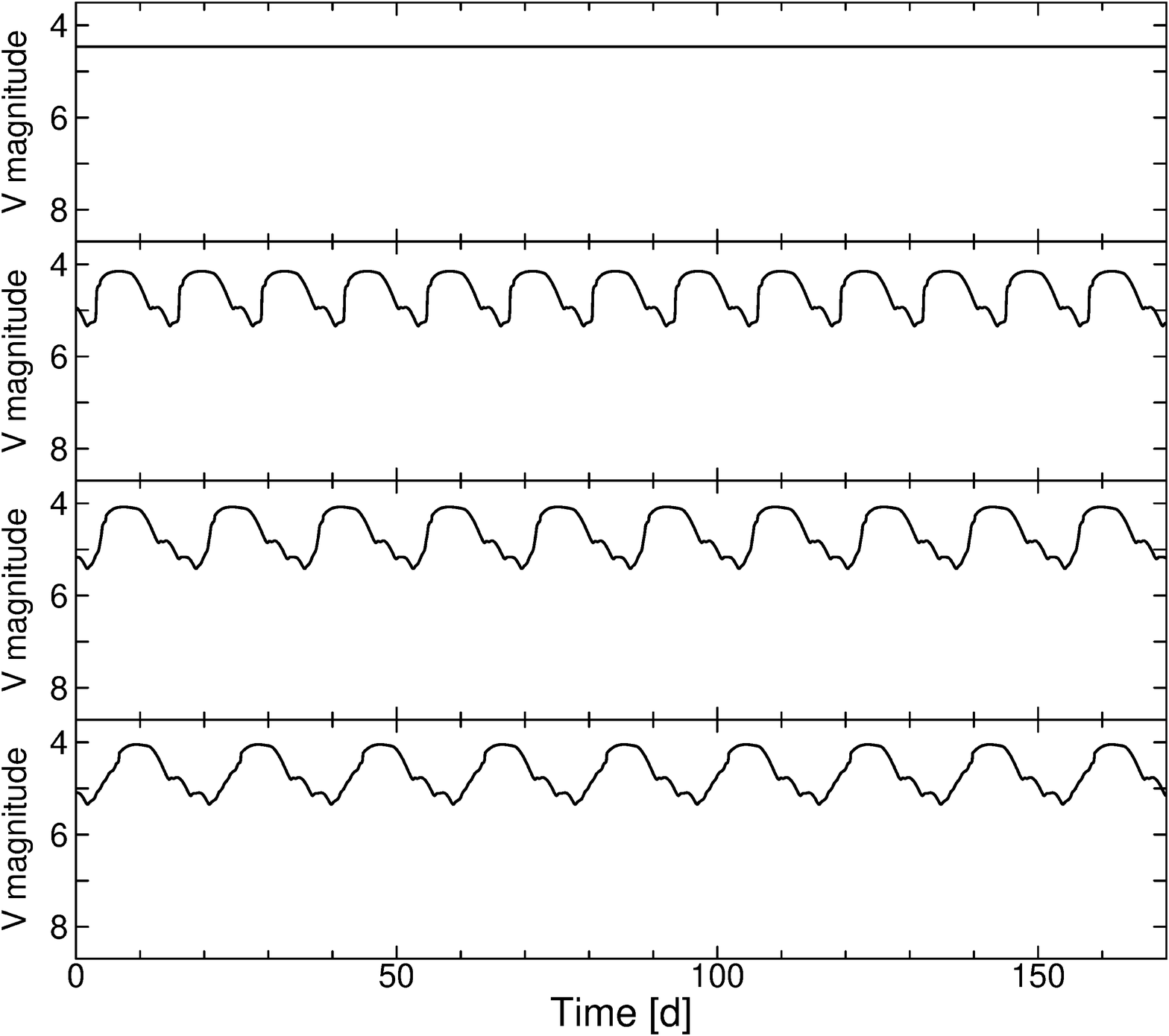}
\end{center}
\caption{Same as Figure \ref{mdot1675vband} but for $\dot{M}_{\rm tr} = 10^{\textcolor{black}{17.25}}$~g~s$^{-1}$ (Models N4, A4, B4, and C4).  }
\label{mdot1725vband}
\end{figure}

In the case of $\dot{M}_{\rm tr} = 10^{16.5}$~g~s$^{-1}$, 
the light curves alternate between several small outbursts and 
a large outburst even in Model C3 just as in Model A1 
when the disk is tilted.  
This is because the cyclic accretion explained in Sec.~5.3 
also occurs and because the mass supply rate in the inner disk 
is not enough to keep it hot even in the highly-tilted disk.  
In the case of $\dot{M}_{\rm tr} = 10^{17.25}$, the disk 
in the standard non-tilted case settles down to the hot 
steady disk, corresponding to NL stars.
In tilted disks, the reduced mass supply in the outer edge 
cannot keep the outer disk hot continuously.  
The outer disk thus occasionally drops from the hot steady state, 
and it produces a light curve with small-amplitude variations.

\section*{4.~~Details for test simulations with another set of binary parameters}

We have confirmed that the main results of our simulations 
do not depend on the binary parameters very much 
by testing with another set of binary parameters in Sec.~5.6.  
Here we present the binary parameters and meshes that we have 
applied to those simulations.  

We have used the binary parameters of KIC 9406652, 
an IW And-type star.  
According to \citet{gie13j1922}, the orbital period 
($P_{\rm orb}$) is 0.2545~d, the white-dwarf mass ($M_{1}$) 
is 0.9$M_{\solar}$, the mass of the secondary ($M_{2}$) is 
0.75$M_{\solar}$, the binary separation ($a$) is 
1.39$\times$10$^{11}$~cm, the tidal truncation radius 
($r_{\rm tidal}$) is 0.328$a$, and the Lubow-Shu radius 
($r_{\rm LS}$) is 0.093$a$, respectively.  
The inner edge of the disk ($r_{0}$) is the same as that 
in our previous calculations.  
We also have changed the number of the meshes 
from 200 to 190, and have given 80, 30, and 80 meshes in the three 
regions defined by equations (A14), (A15), and (A16) 
in Appendix 1.  
Then $r_{\rm input, min}$ is estimated to be 0.053$a$ 
\citep{lub75AD}.


\newcommand{\noop}[1]{}

\end{document}